\documentclass[useAMS,usenatbib]{mnras}
\pdfoutput=1
\usepackage[english]{babel}
\usepackage{graphicx,booktabs}
\usepackage{blindtext}
\usepackage{xcolor}
\usepackage{amssymb,amsmath, wasysym}
\usepackage{tikz}
\usetikzlibrary{backgrounds}
\usetikzlibrary{arrows}
\usetikzlibrary{shapes}

\usepackage{xspace}
\usepackage{siunitx}

\def\apj{ApJ}%
\def\apjl{ApJ}%
\def\apjs{ApJS}%
\def\ao{Appl.~Opt.}%
\def\aap{A\&A}%
\def\mnras{MNRAS}%
\def\prd{Phys.~Rev.~D}%
\def\nat{Nature}%
\def\physrep{Phys.~Rep.}%
\def\jcap{JCAP}

\newcommand{\fv}{F5\xspace}
\renewcommand{\fs}{$|\overline{f_{\textrm{R0}}}| = 10^{-6}$\xspace}

\newcommand{\lcdm}{$\Lambda$CDM\xspace}
\newcommand{\ls}{$768\,h^{-1}\textrm{Mpc}$\xspace}
\newcommand{\lb}{$1536\,h^{-1}\textrm{Mpc}$\xspace}

\newcommand{\mgg}{\textsc{mg-gadget}\xspace}
\newcommand{\fr}{$f(R)$-gravity\xspace}
\newcommand{\ms}{{\rm M}_\odot}
\newcommand{\hompc}{\,h {\rm Mpc}^{-1}}

\begin{document}

\title[MG lightcone project: matter and halo statistics]
{The modified gravity lightcone simulation project I: \\ Statistics of matter and halo distributions}

\author[C. Arnold et al.]
{Christian Arnold,\hspace{-.25em}$^1$\thanks{E-mail: christian.arnold@durham.ac.uk}
Pablo Fosalba,\hspace{-.25em}$^{2,3}$
Volker Springel,\hspace{-.25em}$^{4,5,6}$
Ewald Puchwein$^{7}$\hspace{-.25em}\newauthor
and Linda Blot$^{2,3}$
\\$^{1}$Institute for Computational Cosmology, Department of Physics, Durham University, South Road, Durham DH1 3LE, UK
\\$^{2}$Institute of Space Sciences (ICE, CSIC), Campus UAB, Carrer de Can Magrans, s/n, 08193 Barcelona, Spain
\\$^3$Institut d'Estudis Espacials de Catalunya (IEEC), Carrer Gran Capit\`a 2-4, 08193 Barcelona, Spain
\\$^4$Heidelberger Institut f{\"u}r Theoretische Studien, Schloss-Wolfsbrunnenweg 35, 69118 Heidelberg, Germany
\\$^5$Zentrum f\"ur Astronomie der Universit\"at Heidelberg, Astronomisches Recheninstitut, M\"{o}nchhofstr. 12-14, 69120 Heidelberg, Germany
\\$^6$Max-Planck-Institut f\"ur Astrophysik, Karl-Schwarzschild-Str. 1, D-85741 Garching, Germany
\\$^7$Kavli Institute for Cosmology Cambridge and Institute of Astronomy, University of Cambridge, \\$\phantom{^4}$Madingley Road, Cambridge CB3 0HA, UK
}

\date{\today}
\maketitle

\begin{abstract}
We introduce a set of four very high resolution cosmological simulations for exploring \fr, with $2048^3$ particles in \ls and \lb simulation boxes, both for a $|\overline{f_{\textrm{R0}}}| = 10^{-5}$ model and a \lcdm comparison universe, making the set the largest simulations of \fr to date. In order to mimic real observations, the simulations include a continuous 2D and 3D lightcone output which is dedicated to study lensing and clustering statistics in modified gravity. In this work, we present a detailed analysis and resolution study for the matter power spectrum in \fr over a wide range of scales. We also analyse the angular matter power spectrum and lensing convergence on the lightcone. In addition, we investigate the impact of modified gravity on the halo mass function, matter and halo auto-correlation functions, linear halo bias and the concentration-mass relation. We find that the impact of \fr is generally larger on smaller scales and increases with decreasing redshift. Comparing our simulations to state-of-the-art hydrodynamical simulations we confirm a degeneracy between \fr and baryonic feedback in the matter power spectrum on small scales, but also find that scales around $k = 1\, h\, {\rm Mpc}^{-1}$ are promising to distinguish both effects. The lensing convergence power spectrum is increased in \fr. Interestingly available numerical fits are in good agreement overall with our simulations for both standard and modified gravity, but tend to overestimate their relative difference on non-linear scales by few percent.
We also find that the halo bias is lower in \fr compared to general relativity, whereas halo concentrations are increased for unscreened halos.
\end{abstract}


\renewcommand{\d}{{\rm d}}

\section{Introduction}
\label{sec:introduction}

The question of the nature of gravity is one of the most profound problems in fundamental physics. Although Einstein's General Relativity (GR) has been confirmed to remarkably high precision on small scales \citep{will2014}, there are very few tests of the theory on cosmological scales. 
Upcoming large scale structure surveys like Euclid \citep{euclid} or LSST \citep{lsst} aim to perform such tests by observing the large scale matter distribution of the Universe. In order to fully explore their capacities it is crucial to obtain a detailed understanding of how possible deviations from GR would alter cosmic structure formation and with that the observable large scale structure of the universe. 

In this work we present a set of high resolution cosmological simulations in \fr \citep{buchdahl1970}, which is a possible alternative to GR. \fr has an impact on structure formation in low density environments through a factor of $4/3$ increased gravitational forces \citep[see e.g.][for a recent review]{joyce2015}. For a suitable choice of parameters it nevertheless still passes local tests of gravity \citep{husa2007} as these increased forces are screened in dense environments through the chameleon screening mechanism \citep{khoury2004}. The theory predicts a speed of gravitational waves which is identical to that of the speed of light \citep{ezquiaga2017} and therefore passes the constraints of \cite{GW2017} making it an ideal theory to explore how the possible deviations from GR mentioned above might be observable in upcoming surveys. 

In addition to
providing insight into what plausible alternatives to GR could lool like, 
\fr can -- among other modified gravity theories -- explain the late time accelerated expansion of the Universe without a cosmological constant $\Lambda$. As the origin of $\Lambda$ is theoretically not well motivated and poorly understood, such modified gravity theories have become a very active field of research \citep{joyce2015, sofa2010, hassan2012, clifton2012}. The predicted gravitational wave speed in many of those theories is nevertheless in tension with recent observational data \citep{GW2017}.

The chameleon mechanism which is essential to  screen the modifications to GR in high density environments induces a highly non-linear behaviour of the equations underlying the theory. Therefore, analytic approaches to cosmic structure formation in \fr are even more limited than for GR. Cosmological simulations in modified gravity can on the other hand fully describe these non-linearities and have therefore become the primary tool to study cosmic structure formation in modified gravity \citep{oyaizu2008, li2012, puchwein2013, llinares2014}.

 Cosmological simulation works on \fr include studies of halo and matter statistics \citep{schmidt2010, zhao2011c, li2011, lombriser2013, puchwein2013, arnold2015, cataneo2016}, the properties of voids \citep{zivick2015, cautun2018}, cluster properties \citep{lombriser2012, lombriser2012b, arnold2014}, redshift space distortions \citep{jennings2012} and velocity dispersions of dark matter halos \citep{schmidt2010, lam2012, lombriser2012b}. Weak gravitational lensing in \fr has been investigated as well \citep{shirasaki2015, shirasaki2017, li2018}.
 Hydrodynamical simulations studied the Sunyaev-Zeldovich effect and the temperature in galaxy clusters \citep{arnold2014, hammami2015} and the Lyman-$\alpha$ forest \citep{arnold2015} in \fr. 
High resolution studies of galaxy clusters \citep{moran2014} and Milky Way-sized halos \citep{arnold2016} have been performed as well employing zoomed simulation techniques. In addition cosmological simulations have been used to calibrate scaling relations for the dynamical mass of galaxy clusters in \fr \citep{mitchell2018} which incorporate the non-linearities introduced by the chameleon screening mechanism. 

In this work we introduce the to date in terms of particle number largest simulations of \fr. Employing $2048^3$ simulation particles in boxes of \ls and \lb side-length we performed a set of four simulations in total for the $|\overline{f_{\textrm{R0}}}| = 10^{-5}$ (\fv) gravity model and a \lcdm cosmology for comparison. Along with several time-slice outputs the simulations feature continuous 2D and 3D lightcone outputs which are dedicated to enable clustering and lensing analysis on the lightcone in \fr at a so far unreached precision. 
Similar studies employing large-box high resolution simulations with a lightcone output for \lcdm cosmologies have been carried out previously by the MICE collaboration \citep{fosalba2015b, crocce2015, fosalba2015a}. 

This paper is the first of a series of papers analysing the simulations. It focusses on very high resolution studies of power spectra and correlation functions of both dark matter and halos, halo mass functions as well as halo concentrations and linear halo bias. Making use of the different simulation box sizes a resolution study for cosmological simulations in \fr is carried out as well. We also present a first result on weak lensing although a more detailed study of weak gravitational lensing will be carried out in future work. 

This paper is structured as follows: In Section \ref{sec:fR_gravity} we introduce the theory of \fr and the chameleon mechanism as well as the \cite{husa2007} model. A brief introduction to the simulation code and the simulations carried out within this project is given in Section \ref{sec:simulation_code}. Section \ref{sec:results} presents our results which are finally discussed and summarised in Section \ref{sec:conclusions}. 

\section{$\lowercase{f}(R)$-gravity}
\label{sec:fR_gravity}

\fr is a widely studied modified gravity model \citep{schmidt2010, li2012, puchwein2013, llinares2014} which allows to explain the late time accelerated expansion of the universe without a cosmological constant $\Lambda$. Given its compatibility with the recently observed speed of gravitational waves \citep{GW2017, ezquiaga2017}, it has also become a very important testbed for deviations from GR. 

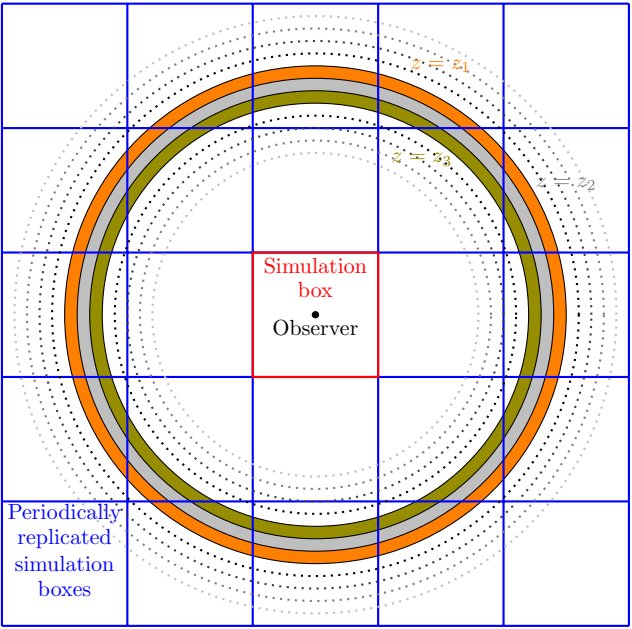
\begin{figure}\centering
\begin{tikzpicture}[scale=0.825, every node/.style={transform shape}]

\draw[fill=orange] (5,5) circle (4cm);
\draw[fill=lightgray] (5,5) circle (3.8cm);
\draw[fill=olive] (5,5) circle (3.6cm);
\draw[fill=white] (5,5) circle (3.4cm);

\draw[fill=black] (5,5) circle (0.05cm);

\draw[step=2cm,thick, blue] (0,0) grid (10,10);
\draw[thick, red] (4,4) rectangle (6,6);

\draw[lightgray, dotted, thick] (5,5) circle (4.8cm);
\draw[gray, dotted, thick] (5,5) circle (4.6cm);
\draw[darkgray, dotted, thick] (5,5) circle (4.4cm);
\draw[black, dotted, thick] (5,5) circle (4.2cm);

\draw[lightgray, dotted, thick] (5,5) circle (2.6cm);
\draw[gray, dotted, thick] (5,5) circle (2.8cm);
\draw[darkgray, dotted, thick] (5,5) circle (3.0cm);
\draw[black, dotted, thick] (5,5) circle (3.2cm);




\node at (5,4.8) {\large Observer};
\node at (5,5.8) {\color{red}\large Simulation};
\node at (5,5.4) { \color{red}\large box};

\node at (1,1.8) {\color{blue}\large Periodically};
\node at (1,1.4) {\color{blue}\large replicated};
\node at (1,1.0) {\color{blue}\large simulation};
\node at (1,0.6) {\color{blue}\large boxes};

\node at (7, 9) {\color{orange}\large $z=z_1$};
\node at (9, 7.1) {\color{gray}\large $z=z_2$};
\node at (6.7, 7.5) {\color{olive}\large $z=z_3$};

\end{tikzpicture}
\caption{Illustration of the method used to construct the full-sky lightcone output. The 400 2D lightcones are equally spaced in lookback-time. For each output, the simulation box is periodically replicated several times in each direction and all particles contained in a thin spherical shell (around an imaginary observer) corresponding to the simulation redshift $z = z_i$ are selected. The selected particles are projected onto a 2D \textsc{healp}ix density map. The thickness $\Delta z$ of the shells is chosen such that they completely cover the volume up to redshift $z = 80$. This approach minimises the repetition of structure in the lightcone. For output times $z<1.4$ the full 3D position information (for the \lb simulations) or a 3D FoF halo catalog (for the \ls simulations) is stored as well.}
\label{fig:lightcone}
\end{figure}
\begin{figure*}
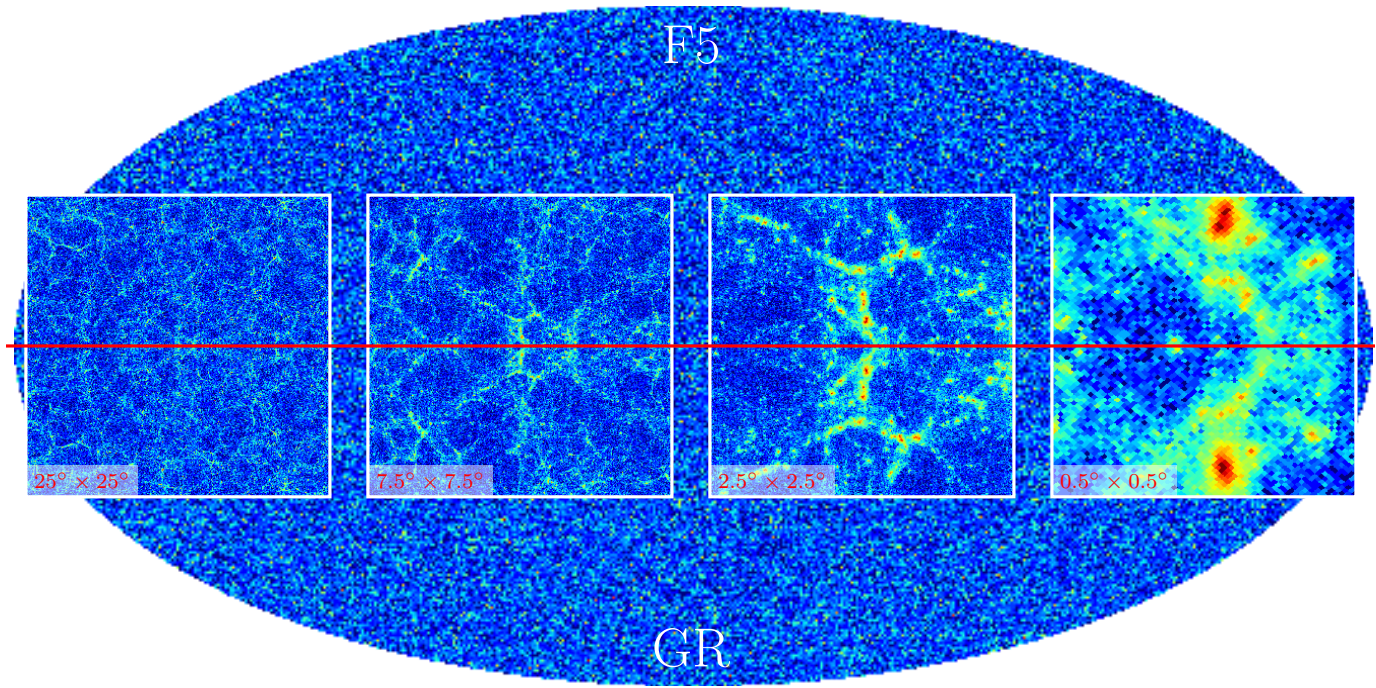

\centering
\include{map_picture}
\caption{A full sky \textsc{healp}ix map in Mollweide projection for $z = 0.5 \pm 0.01$ from the \ls simulation box produced by stacking the 2D lightcone output for the given redshift interval. The upper half of the map shows the \fv simulation output, the lower half the output from the simulation in \lcdm cosmology. The squared maps are zoomed visualisations of the central region of the maps in both theories with a side-length of $25\degr$, $7.5\degr$, $2.5\degr$ and $0.5\degr$.}
\label{fig:map}
\end{figure*}

\fr is an extension of GR. It is constructed by adding a scalar function $f(R)$ to the Ricci scalar $R$ in the action of standard gravity \cite{buchdahl1970}:
\begin{align}
S=\int {\rm d}^4x\, \sqrt{-g} \left[ \frac{R+f(R)}{16\pi G} +\mathcal{L}_m \right],\label{action}
\end{align}
where $g$ is the determinant of the metric $g_{\mu\nu}$, $G$ is the gravitational constant and $\mathcal{L}_m$ is the Lagrangian of the matter fields. 
By varying the action with respect to the metric one obtains the field equations of (metric) \fr, the so called \textit{Modified Einstein Equations}:
\begin{align} G_{\mu\nu} + f_R R_{\mu\nu}-\left( \frac{f}{2}-\Box f_R\right) g_{\mu\nu} - \nabla_\mu \nabla_\nu f_R = 8\pi G T_{\mu\nu} \label{Eequn}.  \end{align} 
The $\nabla$ signs denote covariant derivatives with respect to the metric, $\Box \equiv \nabla_\nu \nabla^\nu$, $T_{\mu\nu}$ is the energy-momentum tensor associated 
with the matter Lagrangian, $R_{\mu\nu}$ is the Ricci tensor and $f_R \equiv \d f(R) / \d R$ is the derivative of the scalar function with respect to the Ricci scalar. 

For cosmological simulations in standard gravity one commonly works in the Newtonian limit of GR, i.e. assumes weak fields and a quasi static behaviour of the matter fields. This assumption is also adopted for most modified gravity simulations (including this work). Its limitations in the context of \fr are discussed in \cite{sawicki2015}.
In the Newtonian limit, the 16 component equation (\ref{Eequn}) simplifies to two equations, a \textit{Modified Poisson Equation}
\begin{align}
 \nabla^2 \Phi = \frac{16\pi G}{3}\delta\rho - \frac{1}{6} \delta R,\label{poisson}
\end{align}
and an equation for the so called scalar degree of freedom $f_R$
\begin{align}
 \nabla^2 f_R =  \frac{1}{3}\left(\delta R -8\pi G\delta\rho\right). \label{fRequn}
\end{align}
$\Phi$ denotes the total gravitational potential, $\delta \rho = \rho - \bar{\rho}$ is the perturbation to the background density $\bar{\rho}$ and $\delta R$ is the perturbation to the background value of the Ricci scalar, i.e. the background curvature. 

In order to simulate cosmic structure formation one has to choose a specific functional form $f(R)$. In order to be consistent with current observational data, the model should 
respect observational limits on deviations 
from GR in our local environment and should lead to a cosmic expansion history which is similar to that in a \lcdm cosmology. For this work we adopt a model which was designed to meet these requirements \citep{husa2007}
\begin{align}
 f(R) = -m^2\frac{c_1\left(\frac{R}{m^2}\right)^n}{c_2\left(\frac{R}{m^2}\right)^n +1},\label{fr}
\end{align}
where $m^2 \equiv \Omega_m H_0^2$ and $c_1, c_2$ and $n$ are parameters of the theory. Throughout this work, we adopt $n=1$. If one sets $\frac{c_1}{c_2} = 6\frac{\Omega_\Lambda}{\Omega_m}$  and $c_2 \frac{R}{m^2} \gg 1,$ the theory closely reproduces the expansion history of a \lcdm universe. In the latter limit, the derivative of equation (\ref{fr}) can be simplified to 
\begin{align} 
f_R=-n\frac{c_1\left(\frac{R}{m^2}\right)^{n-1}}{\left[c_2\left(\frac{R}{m^2}\right)^n+1\right]^2}\approx-n\frac{c_1}{c_2^2}\left(\frac{m^2}{R}\right)^{n+1}.\label{fR}
\end{align}
The remaining free parameter of the theory is now fully described by the background value of the scalar field $f_{R}$ at redshift $z=0$, $\bar{f}_{R0}$. With a suitable choice of this parameter \fr recovers GR in high density regions which is necessary to be consistent with solar system tests through the associated chameleon mechanism \citep{husa2007}. An overview of current constraints on $\bar{f}_{R0}$ can be found in \cite{terukina2014}. Within this work we adopt \fv, which  is in slight tension with local constraints unless there is significant environmental screening by the local group. As we aim to test gravity on much larger scales it is nevertheless still a valuable model to study. Given its slightly stronger deviation from GR compared to models which fully satisfy solar system constraints (such as \fs) it can lead to important insights into how the deviations affect large scale cosmological measures such as weak lensing and clustering statistics. In order to fully explore the GR-testing capacities of upcoming large scale structure surveys like Euclid \citep{euclid} or LSST \citep{lsst} it is critically important to gain a detailed understanding of how these measures are altered by possible modifications to gravity.
\section{Simulations and methods}
\label{sec:simulation_code}

Employing the cosmological simulation code \mgg \citep{puchwein2013} we carry out a set of four collisionless cosmological simulations. Each of the simulations run once for the \fv model and once for a \lcdm cosmology using identical initial conditions. The first pair of simulations contains $2048^3$ simulation particles in a \lb sidelength simulation box, the second pair has the same number of particles in a \ls sidelength box, reaching mass resolutions of $M_{\rm part} = 4.5 \times 10^9\,h^{-1}\ms$ and $M_{\rm part} = 3.6 \times 10^{10}\,h^{-1}\ms$, respectively. All of the runs use the \cite{planck2016} cosmology with $\Omega_m = 0.3089$, $\Omega_\Lambda = 0.6911$, $\Omega_B = 0.0486$, $ h = 0.6774$, $\sigma_8 = 0.8159$ and $n_{\rm s} = 0.9667$. 

\mgg is based on the cosmological simulation code \textsc{p-gadget3}. It is capable of running both hydrodynamical and collisionless simulations in the \cite{husa2007} \fr model. For the simulations presented in this work we use the local timestepping scheme for modified gravity which is described in detail in \cite{arnold2016}. In the following we will give a brief overview of the functionality of the code (a more comprehensive description is given in \citealt{puchwein2013}). 

In order to solve equation (\ref{fRequn}) for the scalar degree of freedom, \mgg uses an iterative Newton-Raphson method with multigrid acceleration on an adaptively refining mesh (AMR grid). To avoid unphysical positive values for $f_R$ which can occur due to numerical values in the simulations, the code solves for $u = \log(f_R / f_{R0})$ instead of computing $f_R$ directly (this trick was first applied by \citealt{oyaizu2008}). Once the solution for $f_R$ is known, one can use it to calculate an effective mass density which accounts for all \fr effects including the chameleon mechanism
\begin{align}
\delta \rho_{\rm eff} = \frac{1}{3} \delta \rho - \frac{1}{24\pi G} \delta R \label{rho_eff}.
\end{align}

By adding this effective density to the real mass density, the total gravitational acceleration can now in principle be obtained using the standard Tree-PM poisson solver which is implemented in \textsc{p-gagdet3}. In order to allow for local timestepping, the standard and the modified gravity accelerations are nevertheless calculated separately for the short range (tree-based) forces (see \citealt{arnold2016} for a more detailed description of the local timestepping in \mgg). 

All four simulations feature a 2D lightcone output consisting of 400 \textsc{healp}ix\footnote{http://healpix.sourceforge.net/} maps \citep{healpix} between redshift $z = 80$ and $z = 0$. The maps are equally spaced in lookback time and have a resolution of $\num{805306368}$ pixels. 
Using the 'Onion Universe' approach \citep{fosalba2008}, they are constructed as follows: If the simulation reaches a redshift $z = z_i$ at which a  lightcone output is desired, the simulation box is repeated several times in all directions such that the whole volume up to the distance corresponding to the redshift $z_i$ around an imaginary observer is covered (see Figure \ref{fig:lightcone}). Subsequently all simulation particles contained in a thin spherical shell around $z_i$ are selected and binned onto the \textsc{healpix} map. The thickness of the shells is chosen such that the lightcone output is space-filling.

Along with the 2D lightcone output the \lb simulation boxes feature a full 3D lightcone output between $z=1.4$ and $z=0$ which is constructed by storing the full 3D position data for all the selected particles within the shell around $z_i$ at a given output time. For the \ls simulations, a 3D halo catalog on the lightcone is produced on the fly instead of the 3D position output.
The centres of the halos are identified using a shrinking sphere approach for all objects identified by the Friends-of-Friends (FOF) halo finder of \textsc{p-gadget3}. Along with their position several properties such as their mass, velocity, center of mass and tensor of inertia are stored.

In addition to the lightcones, the simulation output features several time-slices as well as halo catalogs obtained with the \textsc{subfind} algorithm \citep{springel2001}.

\section{Results}
\label{sec:results}
\begin{figure*}
\centerline{\includegraphics[width=\linewidth]{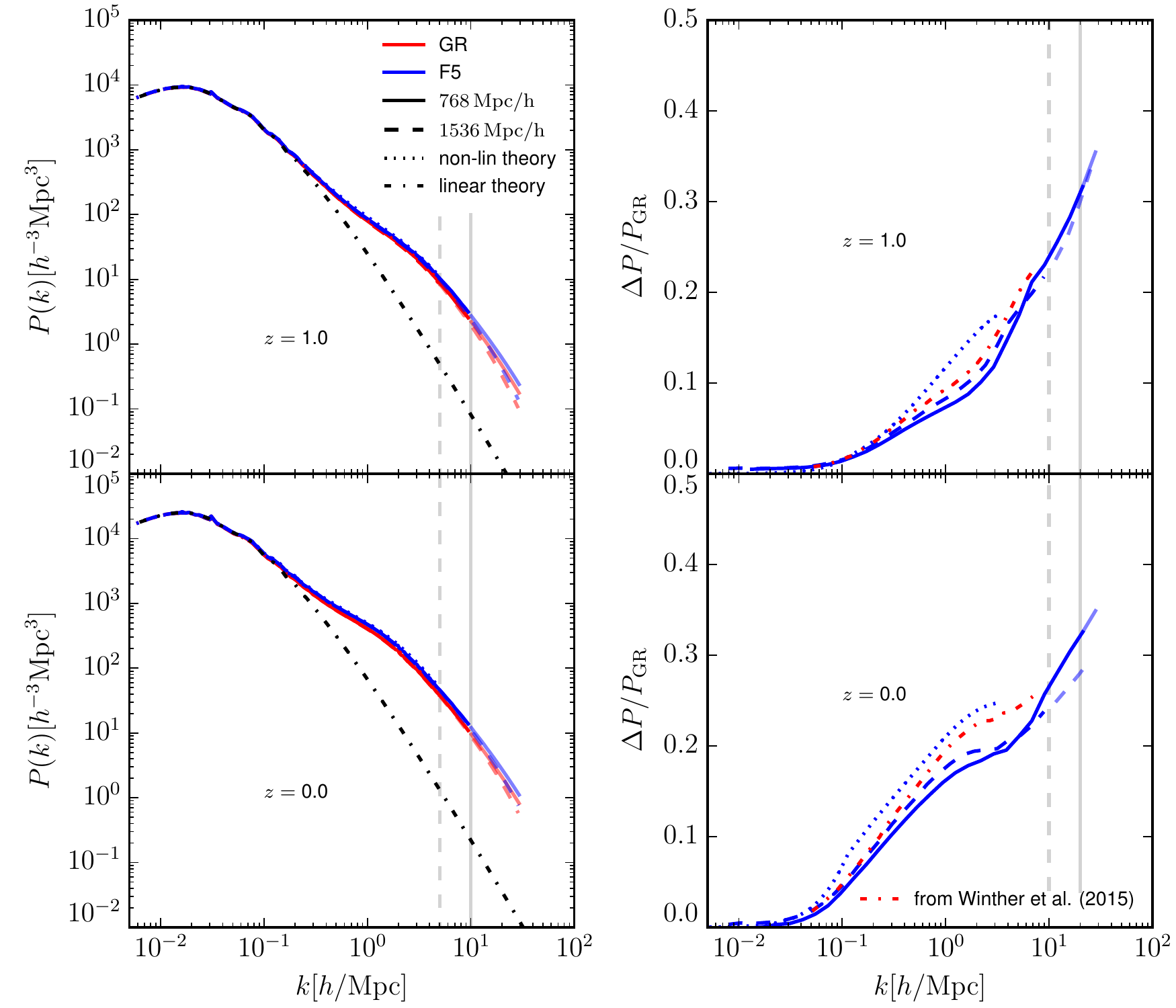}} 
\caption{The matter power spectrum at $z=1$ (\textit{top panels}) and $z=0$ (\textit{bottom panels}) for a \lcdm (\textit{red}) and a \fv universe (\textit{blue}). \textit{Solid lines} refer to results of the \ls simulations, \textit{dashed lines} to the \lb simulation boxes. Numerically converged results are shown as \textit{vivid lines}, results shown as \textit{faint lines} might be affected by resolution. The resolution limits are also indicated by the \textit{grey vertical dashed} and \textit{solid lines} for the large and small simulation boxes, respectively. The dotted lines show non-linear theory predictions from \textsc{halofit} \protect\citep{takahashi2012} and \textsc{mg-halofit} \protect\citep{zhao2014} for standard gravity and an $f(R)$ universe, respectively. Dash-dotted lines display the linearly evolved initial power spectrum. The \textit{right panels} display the relative difference of the \fv simulations to the \lcdm reference simulations. The dotted lines indicate the relative difference predicted by \textsc{halofit} and \textsc{mg-halofit}.  The \textit{dash-dotted orange lines} show the results from \protect\cite{winther2015}.}
\label{fig:matter_powerspec}
\end{figure*}

\begin{figure}
\centerline{\includegraphics[width=\linewidth]{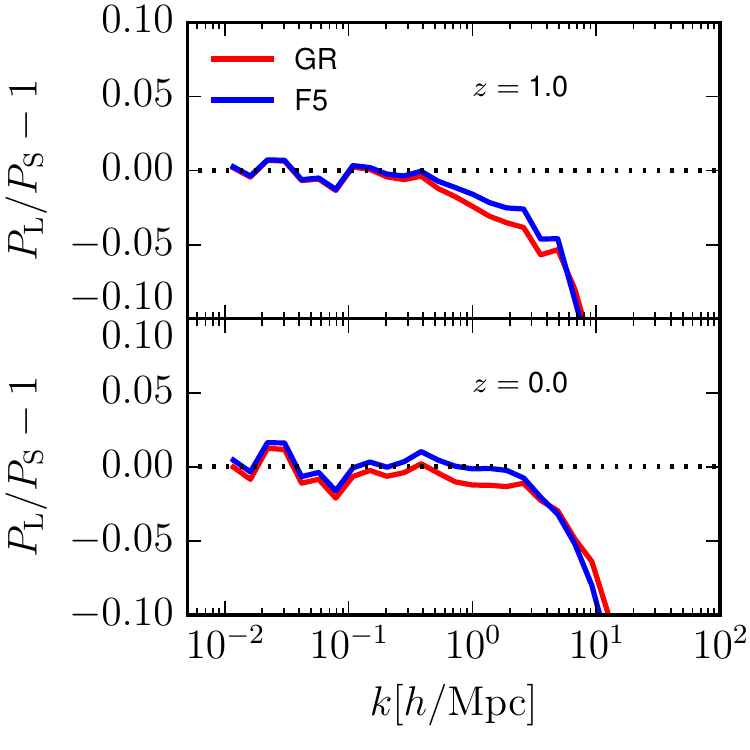}} 
\caption{The relative difference of the power spectra in the \lb simulation box ($P_{\rm{L}}$) with respect to the power spectra in the \ls simulation box ($P_{\rm{S}}$) for a \lcdm cosmology (\textit{red lines}) and \fv (\textit{blue lines}). The \textit{upper panel} shows the comparison for $z=1$, the \textit{lower panel} for $z=0$. \textit{Dotted black lines} indicate equality.}
\label{fig:matter_powerspec_resolution}
\end{figure}


\begin{figure*}
\centerline{\includegraphics[width=0.8\linewidth]{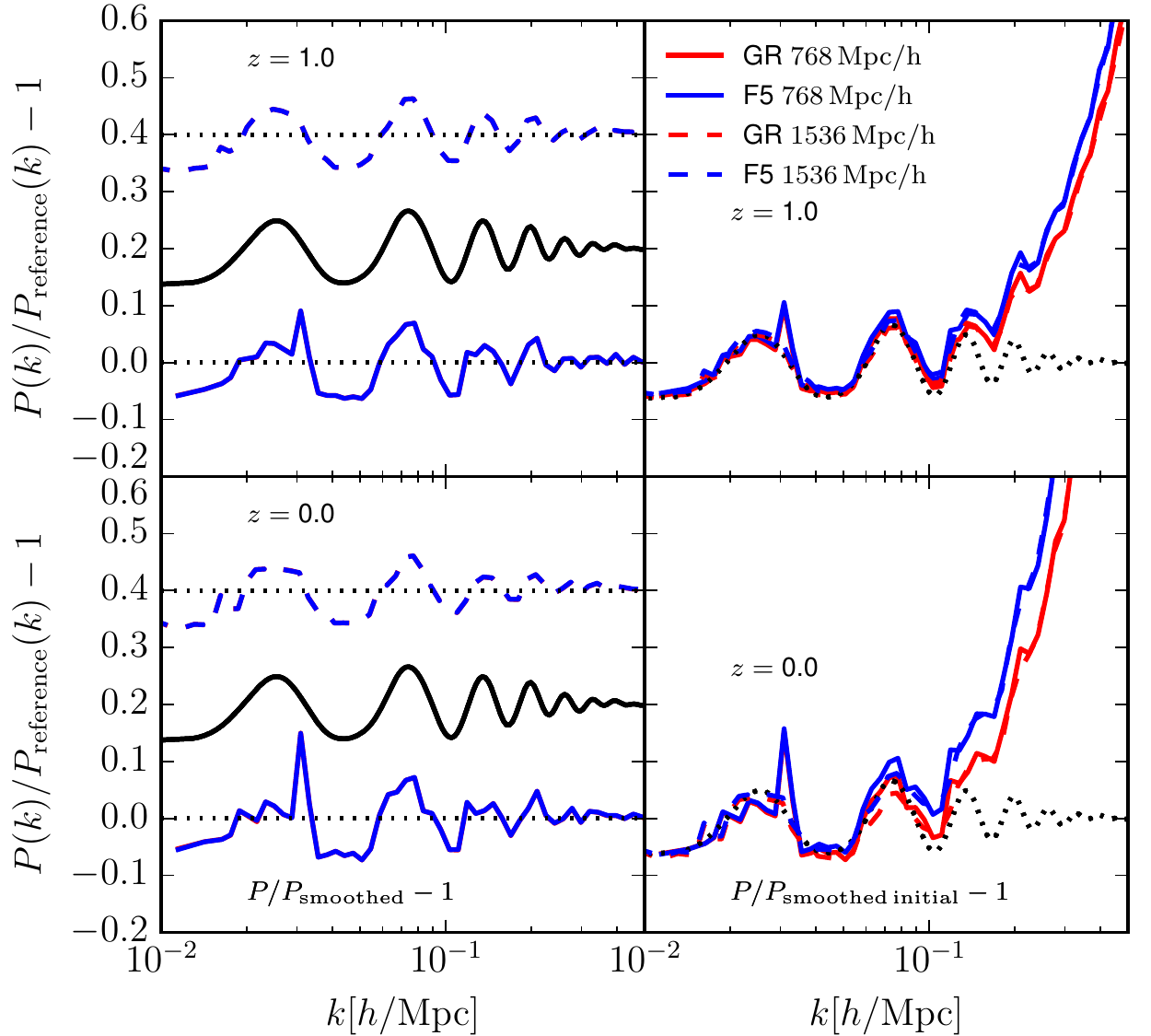}} 
\caption{The relative difference of the matter power spectra in \lcdm (\textit{red lines}) and \fv (\textit{blue lines}) to three different reference power spectra for $z=1$ (\textit{top panels}) and $z = 0$ (\textit{bottom panels}) at the BAO scale. \textit{Solid lines} show results from the \ls simulation box, \textit{dashed lines} results from the \lb simulation box. The \textit{left panels} display the matter power spectrum with respect to the smoothed matter power spectrum. The results for the large box have been shifted vertically for clarity.  \textit{Dotted black lines} represent equality. The \textit{solid black line} indicates the position of the BAO oscillations in the initial power spectrum. Note that the results for both cosmologies are identical in the \textit{left panels} of the plot. The \textit{red lines} are therefore hidden behind the \textit{blue lines}. The \textit{right panels} display the difference of the power spectrum to the smoothed, linearly evolved initial power spectrum. The \textit{dotted black lines} again indicate the BAO oscillations in the linearly evolved initial power spectrum. }
\label{fig:matter_powerspec_wiggles}
\end{figure*}

\begin{figure}
\centerline{\includegraphics[width=\linewidth]{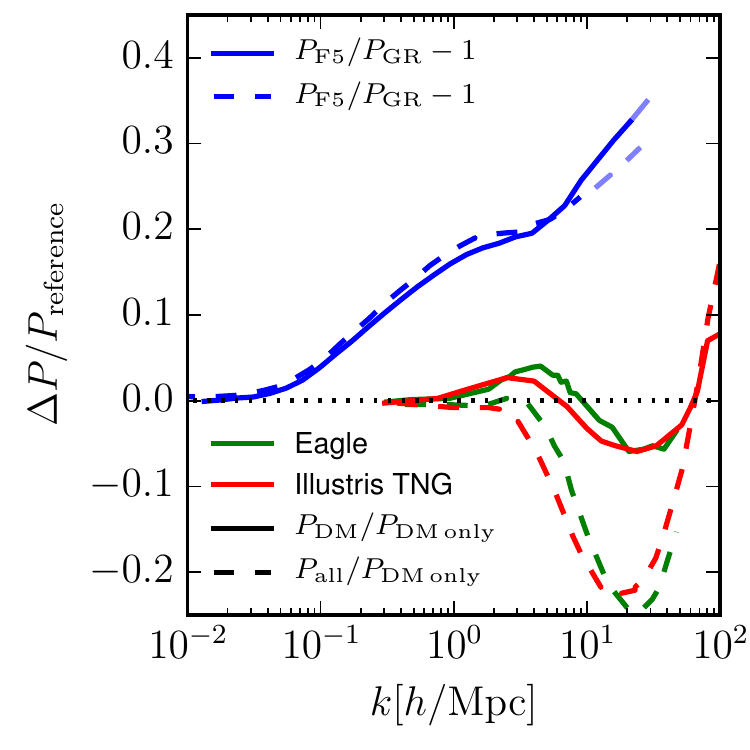}} 
\caption{The impact of different processes on the matter power spectrum at $z = 0$. The \textit{blue lines} show the relative difference due to $f(R)$-gravity explored in this work. The effect of baryonic physics on the dark matter power spectrum in the Eagle (\textit{green}, \citealt{schaye2015}) and Illustris TNG (\textit{red}, \citealt{springel2018}) simulations are indicated by the \textit{solid lines}. The impact of baryons on the total matter power spectrum is indicated by the \textit{green} and \textit{red dashed lines}, respectively. The \textit{dotted black line} indicates equalitiy.}
\label{fig:matter_powerspec_baryons}
\end{figure}

\begin{figure*}
\centerline{\includegraphics[width=\linewidth]{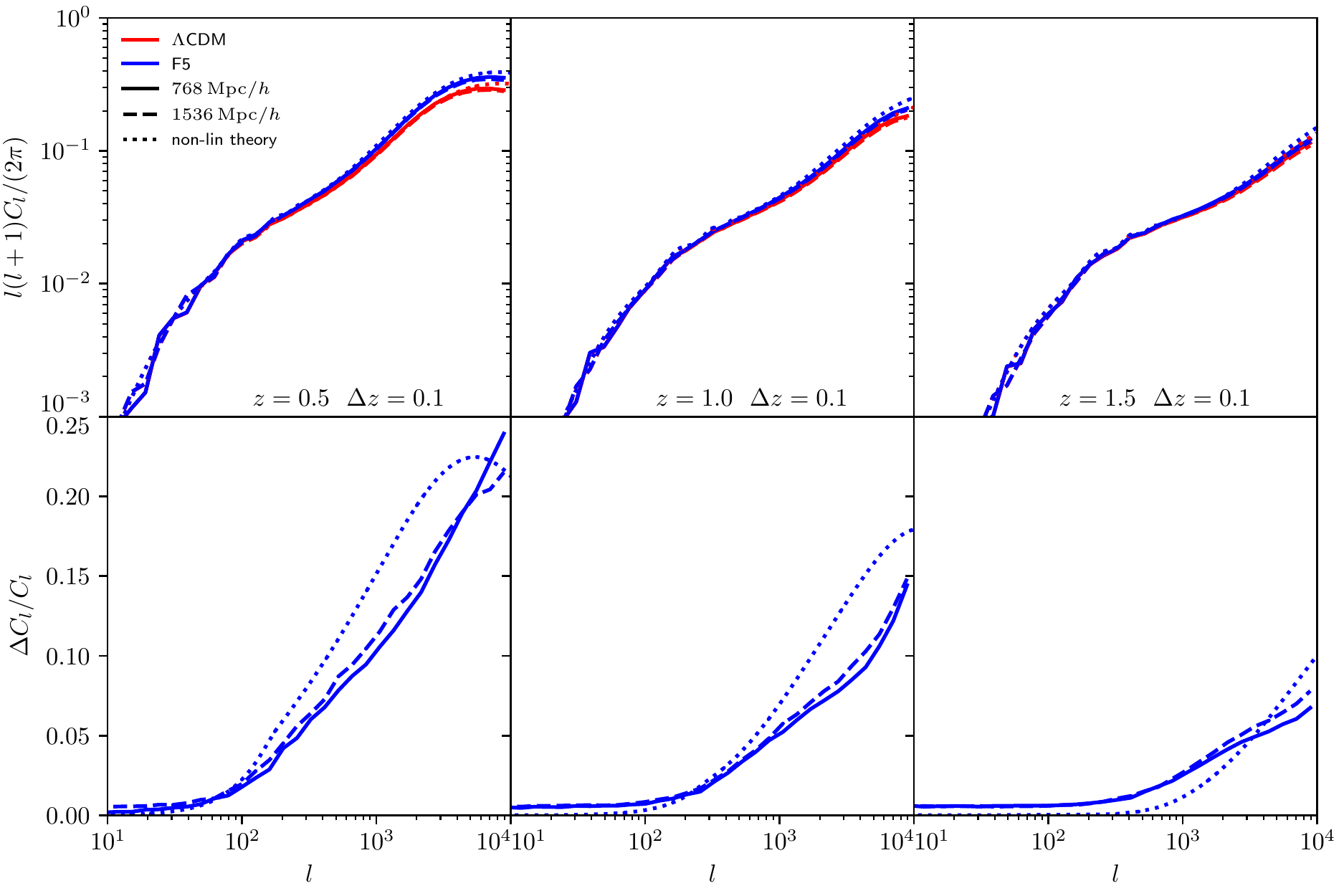}} 
\caption{\textit{Top panels}: The angular power spectrum for the \ls (\textit{solid lines}) and the \lb (\textit{dashed lines}) simulation boxes for a \lcdm and a \fv cosmology at $z = 0.5, 1$ and $1.5$ (\textit{from left to right}). Non-linear theory predictions were derived using \textsc{halofit} \protect\citep{takahashi2012} and \textsc{mghalofit} \protect\cite{zhao2014} and are shown as the \textit{dotted lines}. The \textit{bottom panels} show the relative difference of the \fv results with respect to a \lcdm universe.} 
\label{fig:angular_powerspec}
\end{figure*}

\begin{figure}
\centerline{\includegraphics[width=\linewidth]{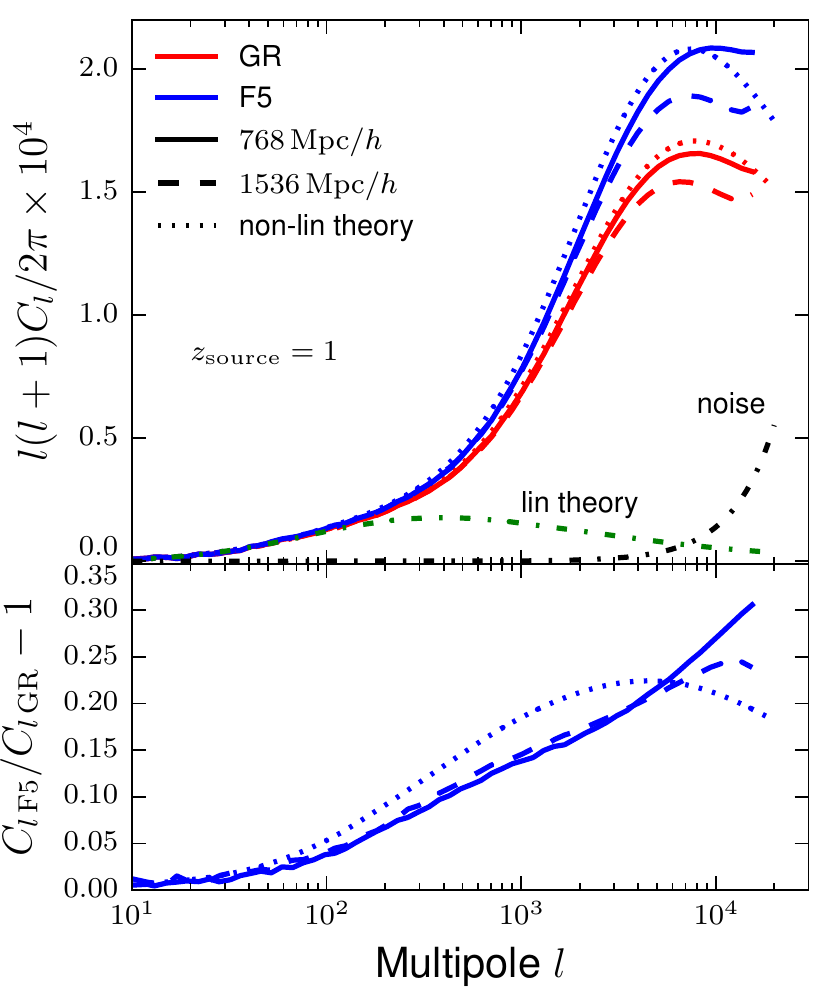}} 
\caption{The lensing convergence angular power spectrum (\textit{top panel}) for sources at redshift $z=0$. \textit{ Red lines} show the results for the \lcdm cosmology simulations, \textit{blue lines} for \fv. The \textit{dashed lines} correspond to the power spectra in the \lb simulation boxes, the \textit{solid lines} to results from the \ls boxes. Non-linear theory predictions were computed using \textsc{halofit} \protect\citep{takahashi2012} and \textsc{mghalofit} \protect\citep{zhao2014} for standard gravity and an $f(R)$ universe, respectively (\textit{dotted lines}). Linear theory predictions are indicated by the \textit{ green dash-dotted line}. The \textit{black dash-dotted line} indicates the shot-noise level for the \lb simulation box. The relative differences between \fr and \lcdm for the two boxes and the non-linear theory predictions are displayed in the bottom panel.}
\label{fig:lensing_convergence}
\end{figure}

\begin{figure*}
\centerline{\includegraphics[width=\linewidth]{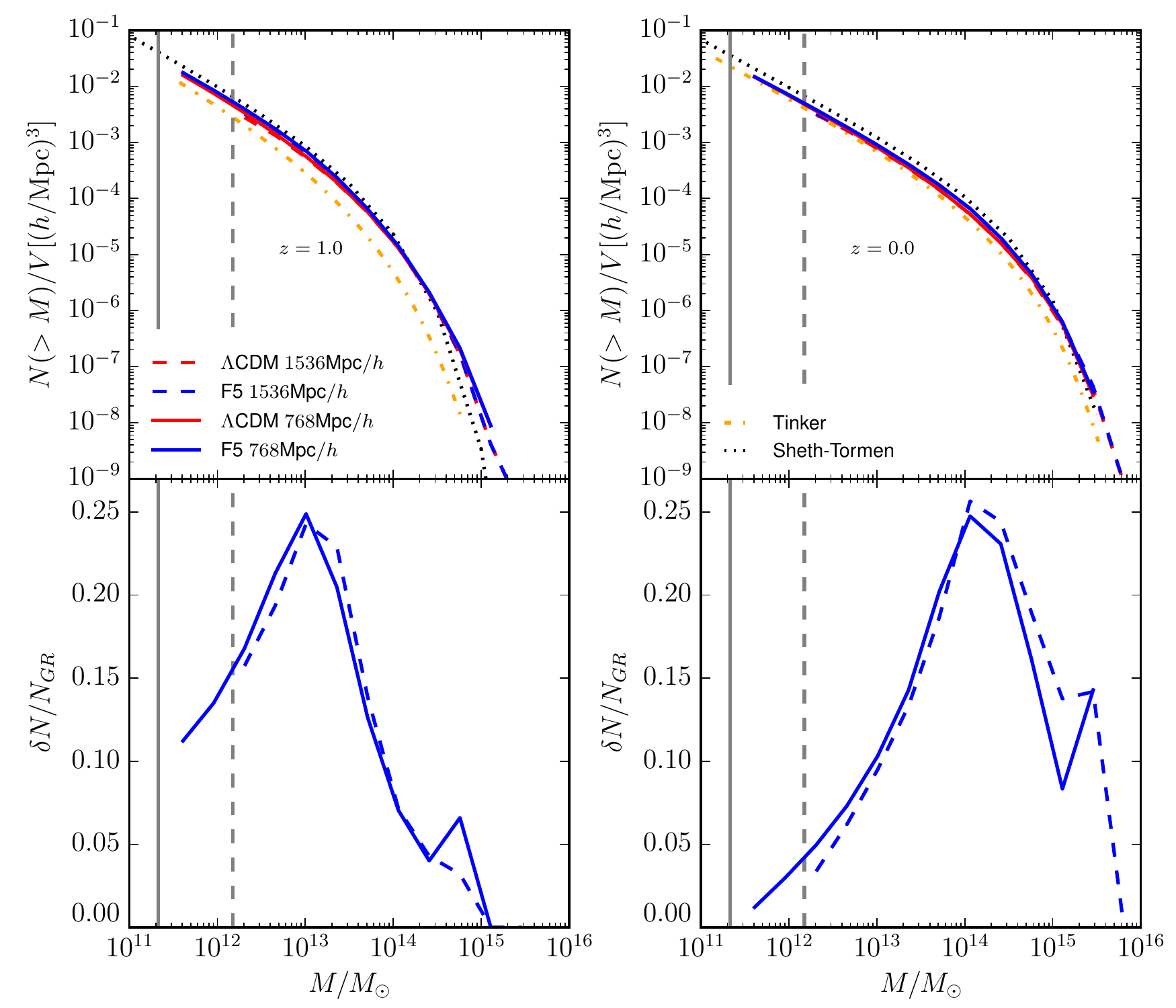}}
\caption{The cumulative dark matter halo mass function at $z=1$ (\textit{left panels}) and $z=0$ (\textit{right panels}). \textit{Solid lines} refer to the \ls simulations, \textit{dotted lines} to the \lb runs. The \textit{bottom panels} show the relative difference between \fv and \lcdm. The vertical \textit{solid} and \textit{dashed} lines indicate the halo resolution limit for the \ls and the \lb simulation box, respectively. Analytical predictions using the methods of  \protect\cite{tinker2010} and \protect\cite{sheth2001} are shown as the \textit{orange dash-dotted lines} and \textit{black dotted lines}, respectively.}
\label{fig:mass_function}
\end{figure*}

\begin{figure*}
\centerline{\includegraphics[width=\linewidth]{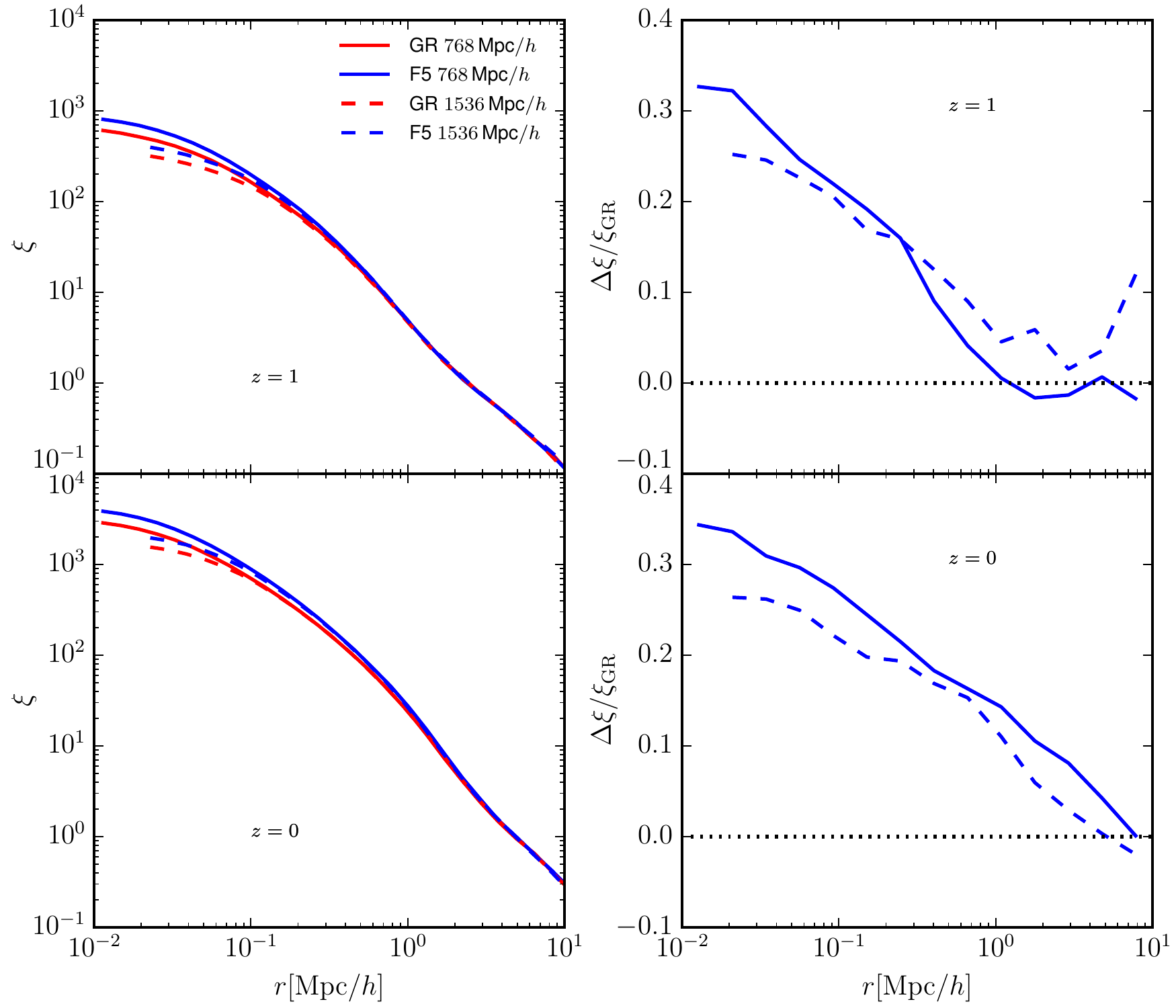}} 
\caption{The dark matter two point correlation function of the two simulation boxes at $z=0$ and $z=1$ for \fv and \lcdm cosmologies. The \textit{solid lines} display results for the \ls simulation boxes, the \textit{dashed lines} for the \lb runs. The \textit{right hand side panels} show the relative difference of \fv to GR.}
\label{fig:correl_dm}
\end{figure*}

\begin{figure*}
\centerline{\includegraphics[width=\linewidth]{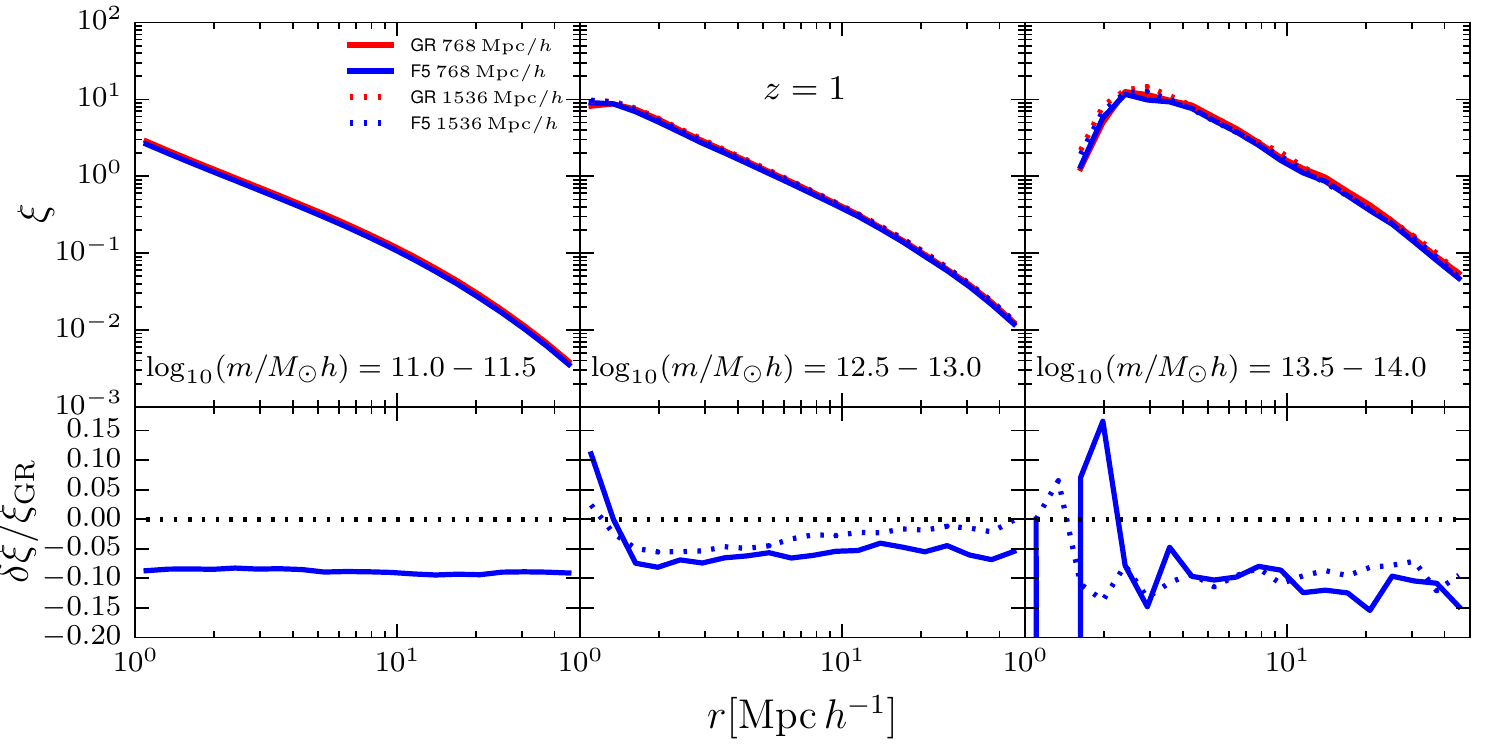}} 
\caption{The dark matter halo two point correlation function at $z=1$ for a \lcdm (\textit{red}) and a \fv (\textit{blue}) cosmology for 3 of the 6 different halo mass ($m_{200\, \rm{crit}}$) bins. The \textit{solid and dotted lines} in the \textit{top panels} refer to the actual halo auto-correlation function for each mass bin for the \ls and \lb simulation boxes, respectively. 
The relative difference in the halo auto-correlation function between a \fv and a \lcdm cosmology is displayed in the \textit{lower panels} for each mass bin, \textit{dotted black lines} indicate zero relative difference. The results from the large box are not shown in the left panels.}
\label{fig:correl_groups_z1}
\end{figure*}

\begin{figure*}
\centerline{\includegraphics[width=\linewidth]{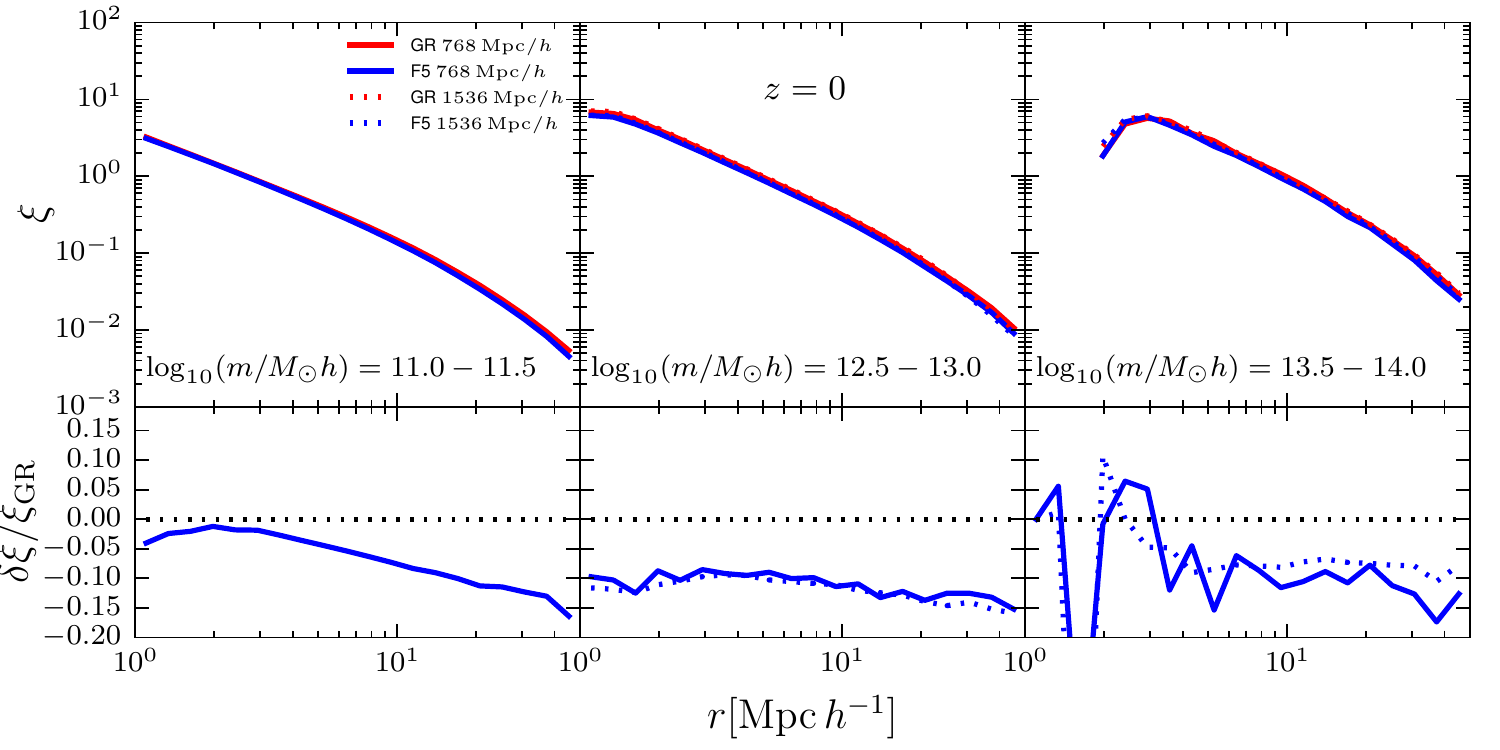}} 
\caption{Same as Figure \ref{fig:correl_groups_z1} but for $z=0$.}
\label{fig:correl_groups_z0}
\end{figure*}

\begin{figure*}
\centerline{\includegraphics[width=\linewidth]{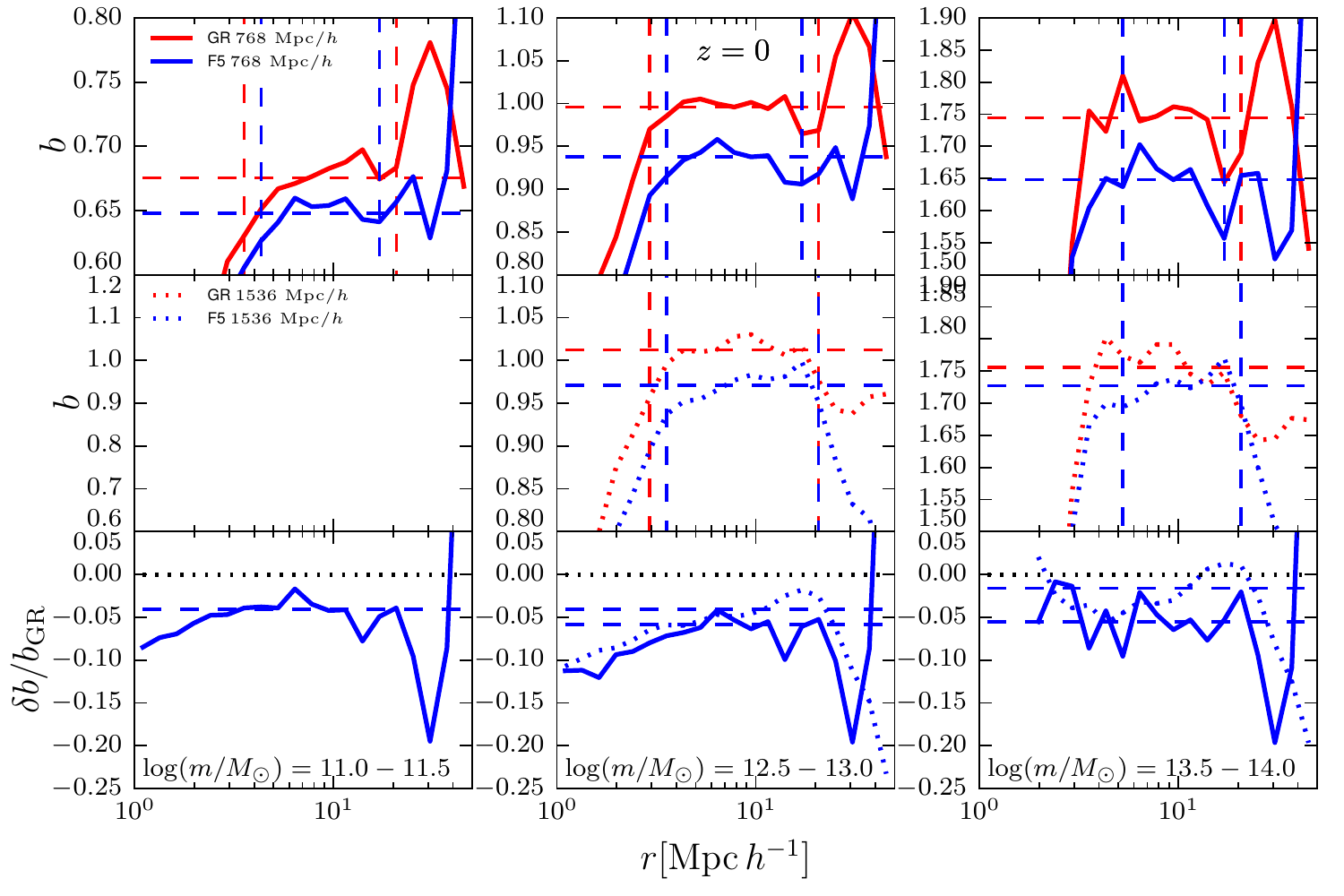}} 
\caption{The halo bias as a function of distance for \lcdm (\textit{red lines}) and \fv (\textit{blue lines}) cosmologies (\textit{top and middle panels}) for 3 of the 6 different mass bins at $z=0$. The results in the \textit{top panels} were obtained from the \ls simulation box (\textit{solid lines}), the results in the \textit{middle panels} from the \lb simulations (\textit{dotted lines}). The \textit{dashed horizontal lines} indicate the median bias, the \textit{dashed vertical lines} show the radial range which is used to infer the median bias. The \textit{lower panels} display the relative difference in the halo bias between \fv and \lcdm, again using \textit{solid} and \textit{dotted} lines for the small and the large boxes, respectively. The \textit{dashed lines} show the difference of the mean, the \textit{black dotted line} indicates zero. }
\label{fig:bias_z0}
\end{figure*}

\begin{figure*}
\centerline{\includegraphics[width=0.9\linewidth]{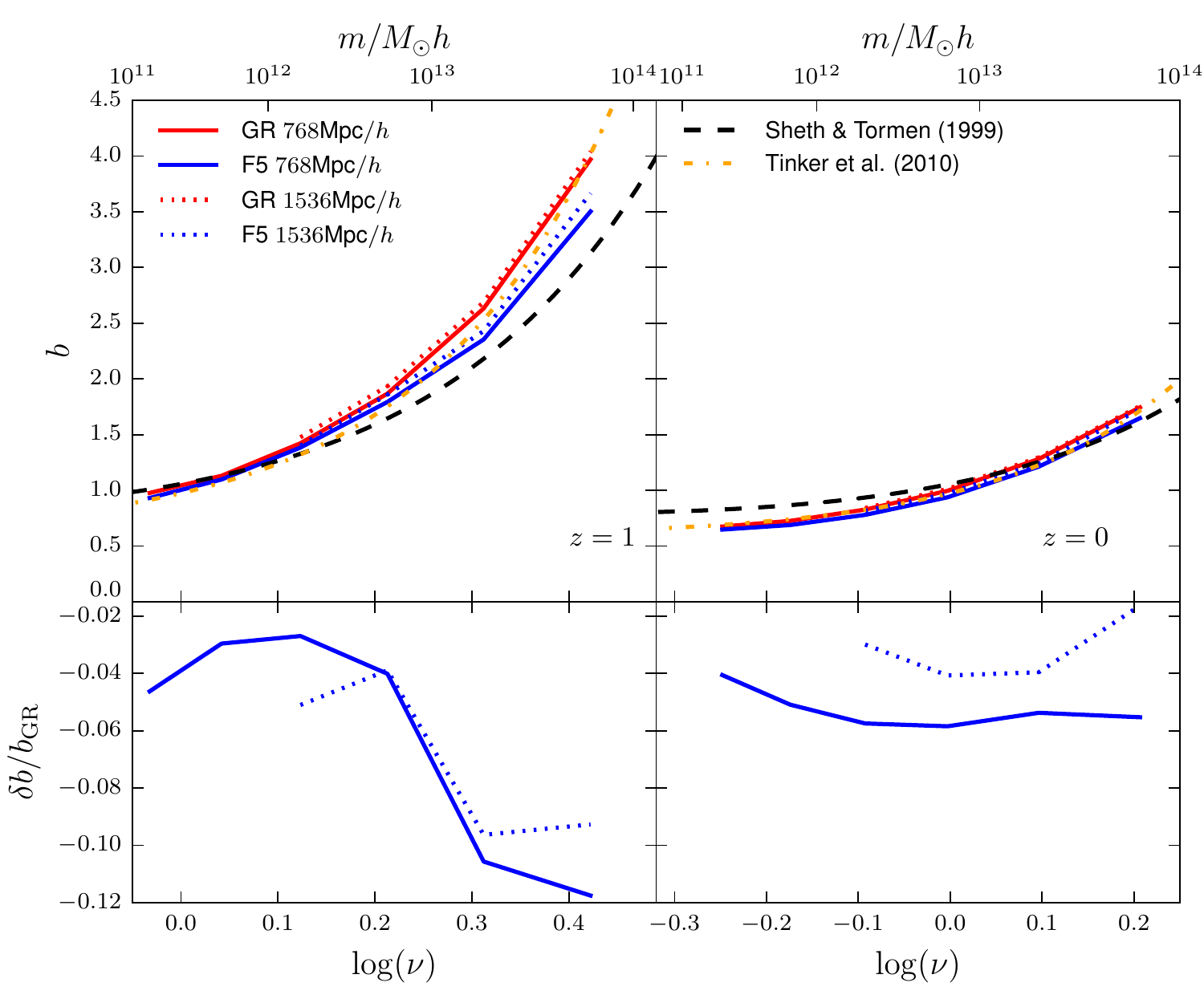}} 
\caption{The mean halo bias as a function of peak height ($\nu$, \textit{lower axis}) and mass (\textit{upper axis}) for \fv (\textit{blue}) and \lcdm (\textit{red}) at $z=1$ (\textit{left panels}) and $z=0$ (\textit{right panels}). The \textit{solid lines} display the results from the small simulation boxes, the \textit{dotted lines} results from the large volume simulations. Analytical predictions from \protect\cite{sheth1999} and \protect\cite{tinker2010} are shown as the \textit{dashed} and \textit{dotted black lines}, respectively. The lower panels show the relative difference between \fv and \lcdm.}
\label{fig:bias_mass}
\end{figure*}

\begin{figure*}
\centerline{\includegraphics[width=\linewidth]{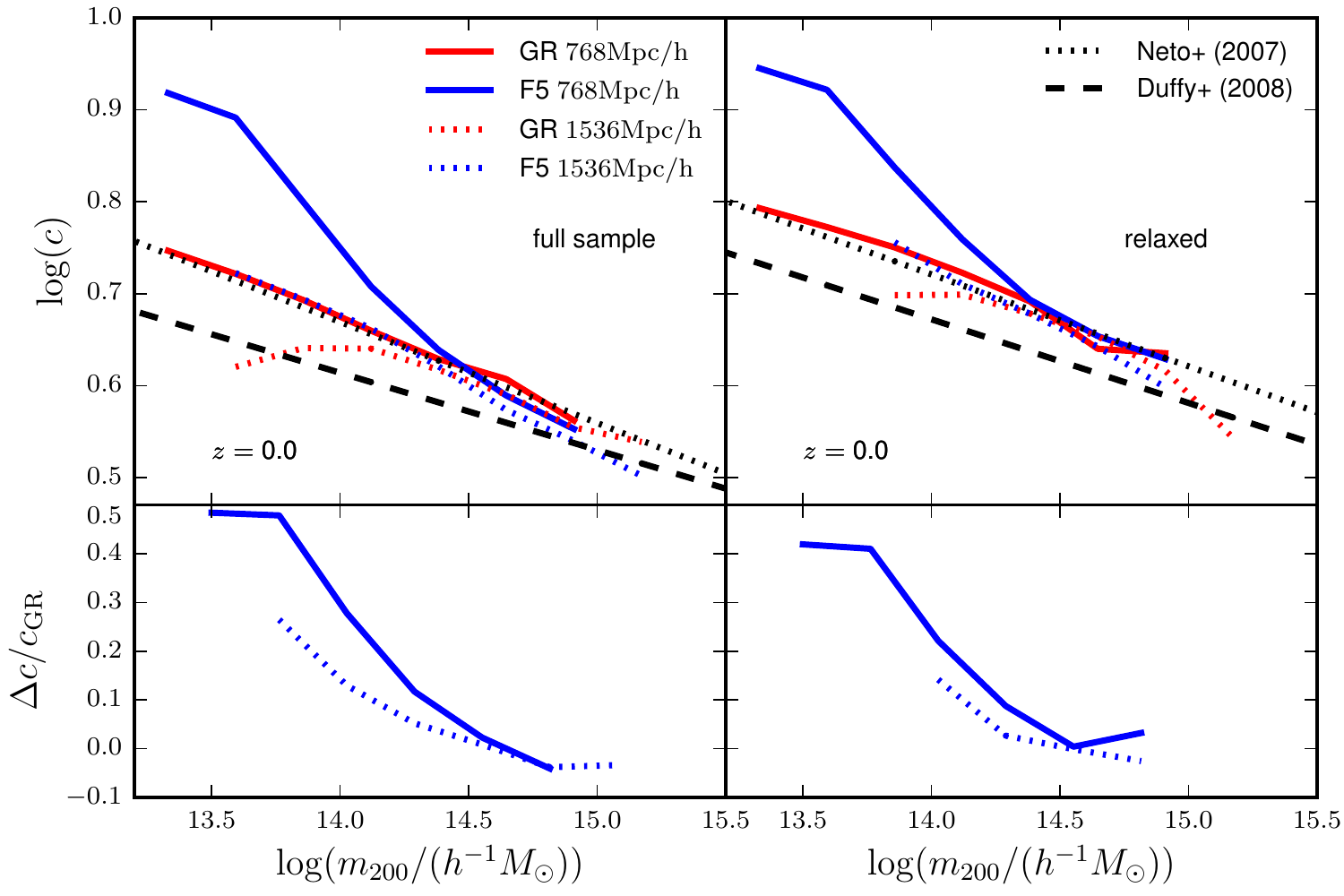}} 
\caption{The concentration mass-relation for \fv (\textit{red}) and \lcdm (\textit{blue}) at  $z=0$. Concentrations for the full halo sample are displayed in the \textit{left panels}, concentrations for the relaxed halos (according to the center-of-mass displacement and subhalo mass fraction criteria described in \protect\citealt{neto2007}) only in the \textit{right panels}. The \textit{dotted lines} display results from the \lb simulation boxes, the \textit{solid lines} for the \ls simulation boxes. The \textit{black dotted} and \textit{dashed lines} show analytical fitting formulae from \protect\cite{neto2007} and \protect\cite{duffy2008}, respectively. The relative difference between \fv and \lcdm is shown in the lower panels. The concentration is calculated from the maxima of the circular velocity curves obtained with \textsc{subfind}. We only plot bins with an error of the median smaller than $0.02$ in $log(c)$.}
\label{fig:concentration}
\end{figure*}

In order to illustrate the 2D lightcone output of the simulations, Figure \ref{fig:map} shows a stacked \textsc{healp}ix density map around redshift $z=0.5$ in Mollweide projection. The map was produced from the \ls simulation box for both the \fv (upper half) and a \lcdm  (lower half) model. Dark blue regions correspond to low matter densities, lighter colours to regions with higher matter densities while the highest densities are indicated by dark red regions.
The squared maps are zoomed projections of the central region of the maps in both cosmological models. 
The maps for the GR model are mirrored along the red line, i.e. they show the same spatial regions as the maps for \fr. One can see from the zooms that the density field on large scales is only mildly altered by \fr while some differences appear on small scales. In order to perform a more quantitative study of this we will consider matter and halo clustering statistics below.
\subsection{Matter and lensing power spectra}

The dark matter power spectrum obtained from the simulations is shown in Figure \ref{fig:matter_powerspec}.  
In order to calculate power spectra over a larger range of scales without performing computationally expensive FFTs for high-resolution grids the density field was folded onto itself twice to obtain the power spectrum at small scales \citep[see][for a more detailed description of the method]{springel2018}. To avoid noise due to the lack of modes at the large scale end of the spectrum, a correction factor for the low-$k$ spectrum was calculated from the initial conditions and used to correct the cosmic variance errors in the power spectra at later times. To ensure that the power spectrum is measured correctly on small scales, we subtract a constant shot-noise correction term from the spectrum. 

The left panels of Figure \ref{fig:matter_powerspec} show the absolute values of the power spectrum at $z=0$ and $z=1$. The right panels give the relative difference of the \fr power spectra with respect to \lcdm. As expected from previous works \fr influences the power spectrum mainly in the regime of non-linear structure growth  \citep{oyaizu2008b, li2013b, puchwein2013, arnold2015}. The relative difference between GR and the \fv model increases with increasing $k$. As the background absolute value of the scalar degree of freedom, $|\bar{f}_R(a)|$, decreases with increasing redshift, one expects a smaller influence of \fr on  the power spectrum at higher redshifts. This is well consistent with the results presented in the plot. At $z=1$, the relative difference reaches about $7\%$ at $k=1\hompc$ but grows to roughly $18\%$ for $z=0$ at the same scale. Note that, although we do not provide statistical errorbars (largely dominated by sample variance in most of the scales shown) for our measurements, we are mainly interested in relative differences or the ratio between F5 and GR, for which sample variance approximately cancels out. 

In order to verify the simulation results, the relative difference in the power spectrum is compared to results from the Modified Gravity Code Comparison Project \citep[][orange dotted lines in the right panels of Figure \ref{fig:matter_powerspec}]{winther2015}. The results are in very good agreement with the relative differences in the power spectra presented in this work. 

In Figure \ref{fig:matter_powerspec_resolution} we present a resolution study for the power spectrum of Figure \ref{fig:matter_powerspec}.  The plot shows the relative difference between the power spectra measured from the large and the small simulation boxes for both the \fr simulations and a \lcdm universe. As one can see from the plot, the spectra of different box-size simulations agree within $~2\%$  up to $k=5\hompc$ for both models at z=0 and within $~4\%$ for $z=1$ and the same $k$-range. We therefore conclude that the absolute value of the matter power spectrum can be trusted up to $k=5\hompc$ for the \lb simulation boxes and up to $k=10\hompc$ for the \ls simulation boxes reflecting the factor of $2$  spatial resolution difference. 
We indicate the range over which we trust the spectrum with the vivid coloured lines in Figure  \ref{fig:matter_powerspec}. The results shown by the faint lines should be treated with caution. 

Although the range where the absolute values of the power spectrum are trustworthy is quite restricted, Figure \ref{fig:matter_powerspec_resolution} shows that both gravity models are affected in a very similar way towards the high-$k$ end of the plot. The relative difference between the modified gravity power spectra and those for the \lcdm models will therefore be reliable until a larger value of $k$ which we estimate to be $k=10\hompc$ for the large simulation box. This conclusion is furthermore supported by the agreement between the results of the \ls and \lb simulation boxes in the right panels of Figure \ref{fig:matter_powerspec} up to $k=10\hompc$. The results from the small boxes can thus be trusted up to $k=20\hompc$. Again, we plot the converged results as vivid lines while results shown as transparent lines might be affected by resolution.

As one can easily see in the right hand panels of Figure \ref{fig:matter_powerspec}, the relative difference in the matter power-spectrum due to the modifications of gravity is consistent between the \lb and the \ls simulation box within the converged range in $k$. The results are still consistent between the two simulations at different resolutions for $k=1\hompc$. At $z=0$ they nevertheless deviate significantly above $k=10\hompc$. This deviation might be caused by an increased, un-physical screening towards the resolution limit of the AMR grid in the simulations. 

To compare our findings for the matter power spectrum to non-linear theory predictions derived with the \textsc{halofit} \citep{takahashi2012} and \textsc{mg-halofit} \citep{zhao2014, hojjati2011, zhao2009} codes for a \lcdm and a \fr universe, respectively, we plot these predictions in Figure \ref{fig:matter_powerspec} as well. The predictions provide a good fit to our simulation data for the relative difference at the low-$k$ end of the plot while small differences appear towards larger values of $k$. 
Note that these predictions have been calibrated on simulation results and thus we choose to show them only where these are well converged  \citep{zhao2014, takahashi2012}.

Figure \ref{fig:matter_powerspec_wiggles} shows the power spectrum at the BAO-scale with respect to different reference power spectra for $z  =1$ and $z = 0$. The panels on the left hand side display the power spectrum divided by the smoothed power spectrum in order to make the BAOs visible. The results for the \lb simulation boxes have been shifted vertically for clarity. The solid black lines (also shifted) show the fluctuations in the initial power spectrum used to create the initial conditions for the simulations linearly evolved to the redshift of the plots. As one can see in the figure, the results for \fr match the results for the \lcdm model very well. There is thus negligible influence of \fr on the growth of the BAO fluctuations. All differences are induced by non-linear structure formation and affect primarily smaller scales. This conclusion is confirmed by the right panels of the plot, showing the power spectrum divided by the linearly evolved initial power spectrum. Note that additional statistical fluctuations in the small-box results due to the lack of large-scale modes with respect to the larger-box simulation.

It has been noted in previous works  that the relative difference in the power spectrum between \fr and \lcdm is of the same order as the relative difference induced by baryonic feedback processes in full-physics hydrodynamical simulations \citep{puchwein2013}. Using the ultra-high resolution simulations performed for this work and state-of-the-art hydrodynamical simulations like those carried out within the Eagle \citep{schaye2015} and Illustris TNG \citep{nelson2018, pillepich2018, springel2018, naiman2018} projects we review this degeneracy in Figure \ref{fig:matter_powerspec_baryons}. We show the relative difference between the \fv simulations and the \lcdm reference runs (blue lines) in comparison to the changes induced by baryons on the total (dashed lines) and the DM (solid lines) power spectrum \citep[the values are from][]{springel2018}. As one can spot from the plot, the relative difference between the DM power spectrum in a full-physics hydrodynamical simulation and a DM-only run is much smaller than the difference due to \fr. The total matter power spectrum is nevertheless suppressed by $20-25\%$ at scales of $k\approx 20\hompc$ due to baryonic feedback. The plot therefore suggests that the effects due to baryons and \fr would approximately cancel at this scale. 

According to the hydrodynamical simulations considered here, the influence of baryons on the power spectrum is on the other hand negligibly small at scales around $k=1\hompc$. There might thus be a sweet spot for testing \fr at these scales with upcoming large scale structure surveys. 
Euclid will e.g. map the dark matter distribution up to $k=5\hompc$ \citep{euclid}, where we measure sizeable deviations $(\approx20\%)$ in F5 models relative to LCDM.
As a cautionary remark we nevertheless have to add that the increased forces in \fr can themselves influence feedback processes. Fully conclusive statements can therefore only be drawn from simulations which include both baryonic feedback and \fr at the same time. One also has to keep in mind that the influence of baryons on the power spectrum is still relatively uncertain \citep{vogelsberger2014, schaye2015, springel2018} and depends on a number of tuneable feedback parameters. It might therefore well be that power spectra of both \fr and \lcdm cosmology can  be brought into agreement with observations by changing the simulation parameters.

The 2D \textsc{healp}ix lightcone output allows us to compute the angular power spectrum at different redshifts. The results are presented in Figure \ref{fig:angular_powerspec}, showing the power at $z = 0.5, 1$ and $1.5$ in the upper panels and the corresponding differences between the spectra in the \fv simulations and \lcdm in the lower panels. The behaviour is similar to the 3D power spectrum. The influence of \fr grows with decreasing redshift. At $z=1.5$, the relative difference reaches roughly $7\%$ at a multipole number of $l   = 10^4$. At the same scale, the relative difference grows to $15\%$ at $z=1$ and further increases to $25\%$ at redshift $0.5$. 
This result is consistent with what we measured for the 3D $P(k)$. In the Limber limit, there is a one to one relation between comoving wavenumbers and multipoles at a given redshift $z$, $l=k \times r(z)$. For instance, for $z=1$, we obtained that F5 exceeded LCDM by $15\%$ at $k=5\hompc$, which projects onto $l\approx10^4$, which is what we observe in the lower central panel of Figure \ref{fig:angular_powerspec}.
At lower multipole numbers the effects due to \fr are smaller. For $z  =0.5$, the increased screening effect due to the lack of resolution at small scales in the \lb simulations is also visible in the angular power spectrum: The large simulation boxes show an approximately $4\%$ lower relative difference at  $l=10^4$ compared to the \ls boxes. 
As for the 	3D matter power spectrum we compare the simulation results to \textsc{halofit} and \textsc{mg-halofit} predictions (Note that these are obtained using the Limber approximation). These provide a very good fit to the absolute values of the angular power on large and intermediate scales as well while small differences appear at high multipoles around $l\approx 10^4$. Their predictive power for the relative difference between \fr and \lcdm universes is nevertheless limited: While \textsc{halofit/mg-halofit} and simulations show reasonable agreement at low $l$, the differences reach up to $5\%$ for larger multipoles. 

Continuing the analysis of the 2D lightcone output we show the weak-lensing convergence power spectrum in Figure \ref{fig:lensing_convergence} for sources at redshift $z=1$. We compare our results from the large and small simulation boxes for both models to linear and non-linear theory predictions for both gravity models in the upper panel of the plot. The relative differences between \fr and the \lcdm simulations for both boxes and the theory predictions are shown in the lower panel. 
The simulation results agree very well between the \ls and the \lb simulation box up to $l=\num{3e3}$. At smaller angular scales (i.e, larger multipoles), the results start to deviate reaching a $5-10\%$ difference at $l=10^4$. 
We thus conclude that the results are converged until $l=\num{3e3}$ and there are mass-resolution effects beyond this scale.

The theoretical predictions for GR have been derived using the \textsc{halofit} package \citep{takahashi2012}. 
The simulation results are in general in good agreement with the \textsc{halofit} predictions.
 Above $l=40$ the fitting formulae are slightly over estimating the lensing convergence power. For the \fv model we used \textsc{mg-halofit} \citep{zhao2014} to derive theoretical predictions. Again, the simulations show a lower lensing convergence power compared to these predictions. A similar discrepancy between \textsc{mg-halofit} and simulations has already been observed by \cite{tessore2015}.

As expected from the linear matter power spectrum the relative differences between the modified gravity model and \lcdm in the lensing convergence are very small at linear scales and increase towards larger multipoles. The results from the large and the small simulation box again agree up to $l=\num{5e3}$ and reach a value of $20\%$ at this scale. 
The main contribution of the 3D dark-matter power to the convergence angular power spectrum for sources at $z=1$ comes from lenses at $z\approx0.5$, where we found (see Figure \ref{fig:matter_powerspec}) that that the F5 model exceeds LCDM by a comparable amount $(15-20\%)$ at the corresponding scale given by the Limber limit relation, $k=l/r(z=0.5) = 5000/1400 \approx 4 \hompc$. 
This result is consistent with the findings of \cite{li2018}. The relative difference between the \mbox{\textsc{(mg-)halofit}} predictions for \fr and standard gravity is approximately $5\%$ larger than the one measured from the simulations in the non-linear regime, and it drops below the simulation result on higher multipoles. 

\subsection{Halo mass function}
The cumulative halo mass-function is shown in the upper panels of Figure \ref{fig:mass_function}. The lower panels show the relative difference between the considered modified gravity model and a \lcdm cosmology. The mass-functions have been normalised by volume in order to make the two different simulation box-sizes directly comparable. The halo resolution limits given by $m_{{\rm halo}} > 32 \times m_{{\rm part}}$ (the minimum number of particles per group identified by \textsc{subfind} is $32$) are indicated in the plot by the solid and dashed vertical grey lines for the small and the large simulation box, respectively.
As expected, the large simulation boxes can not form low mass halos due to a lack of resolution. The curves for the \lb boxes therefore do not reach the low mass end of the plot. The \ls simulations on the other hand can not form halos with masses above $\num{3e15} \ms$ because of the limited volume. 

The mass functions are enhanced in \fr with respect to GR. The relative difference between the models reaches $25\%$ at $\approx 10^{13}\ms$ for redshift $z=1$. The difference decreases towards lower and higher masses. At $z=0$, the relative difference has a maximum at $\approx 10^{14} \ms$. This behaviour is consistent with what one would expect from the evolution of the background value of the scalar field $\bar{f}_R(a)$. 

At high redshift, the background value of the scalar field is smaller \citep[see, e.g.][]{arnold2014}. The mass threshold for screening is therefore lower and \fr mainly affects lower mass halos. These halos will consequently grow faster and become more massive leading to more intermediate-mass halos in the mass function, compared to GR. 
Towards lower redshift, the mass threshold for screening shifts towards higher masses while the intermediate-mass halos at the same time continue growing faster than in GR. The peak in the mass function will consequently shift towards higher masses with decreasing redshift. 

The results for the halo mass function are also consistent with those found in \cite{winther2015}. Both the relative differences found in this work and in \cite{winther2015} are smaller than those reported in \cite{schmidt2009}, who considered the mass function in \fr using $M_{300\, {\rm crit}}$ as opposed to  $M_{200\, {\rm crit}}$ which is used in this work. As the density in the central part of the halos is higher in \fr compared to a \lcdm model \citep{arnold2016}, a larger difference in the mass function using $M_{300\, {\rm crit}}$ is reasonable. 
It is worth noting that the simulations employed in this work have both better mass resolution and larger box-sizes compared to these previous works. We can therefore analyse the mass function over a much wider range of scales and with significantly better statistics.  

Comparing the relative difference between the analytical fitting functions to the difference between the modified gravity model and GR simulations it is obvious that the theoretical uncertainties are much bigger than those induced by the gravity model. The gap between theoretical predictions and simulations is much  larger than the difference between the gravitational theories themselves. 
The $<25\%$ change in halo abundance, corresponds to only a small shift on the mass axis. Given uncertainties in halo mass measurements, it thus seems very challenging to use halo mass functions for constraining the deviations from GR considered here.

\subsection{Matter and halo correlation functions}
Figure \ref{fig:correl_dm} shows the DM two point correlation function for the four simulations at redshift $z=1$ and $z=0$.  The correlation functions are calculated in real space employing the gravity tree of \mgg. The right hand side panels display the relative difference between the \fv model and a \lcdm universe. As for the quantities considered above, the impact of \fr is larger on small scales. At $z =0$, the relative difference reaches about $35\%$ at scales of $r = 10^{-2}\,h^{-1}{\rm Mpc}$, decreases roughly linearly in $\log(r)$ and reaches zero at $r\approx 10{\rm Mpc}/h$. For redshift $z=1$, the relative difference due to modified gravity at small scales is approximately the same. It nevertheless decreases faster towards large scales reaching $\Delta \xi / \xi_{\rm{GR}} =0$ at $ r \approx 1 {\rm Mpc}/h$.
The relative difference between \fr and GR is smaller for the large simulation boxes at small scales. This effect is caused by the unphysical flattening of the correlation functions once the spatial resolution limit of the simulations is approached (we use a gravity softening of $0.01 h^{-1}{\rm Mpc}$ and $0.02 h^{-1}{\rm Mpc}$ for the small and the large simulation boxes, respectively). 

The halo-halo twopoint correlation functions were analysed by splitting the halo sample identified by \textsc{subfind} into six different $m_{200\, \rm crit}$ mass-bins for all four simulations performed within this project. The mass bins are selected such that they span at least $0.5$ dex in mass, but also contain at least $10^5$ halos to ensure sufficiently low noise. The resulting bin-boundaries are $\log(m \ms^{-1} h^{-1}) = 11-11.5, 11.5-12, 12-12.5, 12.5-13, 13-13.5, 13.5-14$. Figure \ref{fig:correl_groups_z1} shows these auto-correlation functions for three of the mass bins at redshift $z=1$. Relative differences between standard and \fr are displayed in the lower panels. 
The results for the intermediate (center panel) and the high mass bin (right panel) shown in the plot are consistent between the \ls and \lb simulation boxes. 
We do not show the results of the large box simulations in the left panel as the two lowest mass-bins are below the halo mass-resolution limit shown in Figure \ref{fig:mass_function}. 
At small radii, the halo two point correlation functions are affected by the finite size of the halos, limiting the minimal distance between two halos of a given mass. This effect is especially pronounced for larger mass halos whose distance is limited by (twice) their radius. The correlation functions in the intermediate and the high mass bin therefore decrease towards lower radii and cannot be used as a meaningful cosmological observable at these length scales. 

The relative difference in correlation between \fr and the \lcdm simulations does not show a strong dependence on radius or mass in Figure \ref{fig:correl_groups_z1}. The halos are about $10\%$ less correlated in \fr compare to GR at both the high and the low mass end of the halo mass function in our simulations, while the relative difference is about $7\%$ in the high-resolution simulation for intermediate masses. The relative difference in correlation functions from the big simulation box slightly decreases with increasing radius in this mass bin, from about $5\%$ at low radii to no significant difference at large radii. This effect is likely caused by the limited mass resolution of the \lb simulation box. We therefore think that the results from the small simulation box are more reliable for this mass bin. 

Figure \ref{fig:correl_groups_z0} displays the same quantities at redshift $z=0$. Halo-halo correlation functions are again shown for three of the mass bins in the upper panels while the corresponding relative differences are plotted in the lower panels. The results from the \lb simulation box are again not shown for the lowest mass bin. 
As for $z=1$, the correlation functions for the other mass bins are consistent between the two independent simulations of different mass resolution. The relative differences between the modified gravity simulations and standard cosmology are of the order of $10\%$. They nevertheless show a slight dependence on radial scale which is most pronounced at low halo masses. The relative difference is approximately zero for low radii in the left panel of the plot and decreases to $-10\%$ at $r\approx 40 \textrm{Mpc}/h$. As the lowest mass bin is at the resolution limit of the small simulation box, this result should nevertheless be taken with caution. Higher mass halos are about $8\%$ less correlated in \fr compared to GR at low radii and show roughly $12\%$ difference at large radii in the plot.

\subsection{Linear halo bias}
We define the linear halo bias as $b(m, z)^2 = \xi_h(m, z, r) / \xi_m(z,r)$, where $\xi_h$ is the halo auto-correlation function shown in Figures  \ref{fig:correl_groups_z1} and \ref{fig:correl_groups_z0}, and $\xi_m$ is the matter auto-correlation function. 
The halo bias is scale dependent on small scales but asymptotically flattens towards larger radii reaching a constant value at large scales ($r \approx 10\, h^{-1} {\rm Mpc}$). This is also visible in the top and middle panels of Figure \ref{fig:bias_z0}. While the bias is strongly scale dependent up to $r \approx 3 - 5 h^{-1} {\rm Mpc}$, depending on halo mass, it becomes constant at larger scales. At the large radius end of the plots in Figure \ref{fig:bias_z0}, the bias is dominated by the noise which occurs in the matter correlation functions at large scales. 
These findings are consistent with the results of \citep[][see Fig. 16]{crocce2015} who find the halo bias to be scale independent at a percent-level for scales larger than $15-20 h^{-1}\,{\rm Mpc}$ in the MICE-GC simulation, with some deviation from scale-independence on smaller scales, depending on halo mass. The degree of scale-dependence found also depends on the estimator used (halo-matter vs. halo-halo correlations).
Here, we are primarily interested in the bias in the constant regime, which is usually referred to as \textit{linear halo bias} \citep{kravtsov1999}.

The procedure to obtain the bias is illustrated  in Figure \ref{fig:bias_z0} for the same halo mass bins shown in the previous figures (again, we do not show results for the large box in the left panels). The top panels show the (scale dependent) bias for the \ls simulation boxes. The same quantity is shown for the large boxes in the middle panels. Bottom panels display relative differences between \fr and a \lcdm cosmology for both box-sizes. In order to obtain the region in which the scale dependent bias is roughly constant, we calculate its mean over all radii and select radial scales where the bias differs by less than $50\%$ as our fitting region (indicated by the vertical dashed lines in the top and middle panels). Our result for the (scale independent) bias is the median (scale dependent) bias in this region (horizontal dashed lines in the top and middle panels). 
The relatively large deviations from the mean bias at large and small radii in the plots are caused by the difficulties in measuring correlation functions at these scales, as discussed before. The results for the intermediate and the high mass bin for the large and the small simulation box are nevertheless consistent, showing that our results are reliable for these masses.

Figure \ref{fig:bias_mass} shows the results for the scale independent halo bias as a function of mass and peak-height parameter $\nu = \delta_c / \sigma (m)$. $\delta_c = 1.686$ is the (linearly estimated) critical over-density for spherical collapse. $\sigma^2 (m)$ denotes the variance of the linearly evolved initial density field at $z=0$ at mass scale $m$. It can be calculated from the convolution of the  linear matter power spectrum with a real-space top hat filter $W_r(k)$ of width $r(m) = \left(\frac{3m}{4\pi \bar{\rho}}\right)^{\frac{1}{3}}$,
\begin{align}
\sigma^2(m) = \frac{1}{(2\pi)^3} \int P(k, z) |W_r(k)|^2 4 \pi k^2 \d k.
\end{align}
The top panels of the plot in Figure \ref{fig:bias_mass} show the bias at redshift $z=1$ (left) and $z=0$ (right) for both simulation boxes and models. We restrict the mass range to the well-resolved regions in the simulations. Theoretical predictions from \cite{sheth1999} and \cite{tinker2010} are shown as the dashed and dotted black lines, respectively. 

For all mass bins, the absolute values of the correlation functions agree very well between the simulation boxes. At $z=1$, the theoretical predictions of \cite{tinker2010} are well reproduced by our simulations for standard gravity. This is as well the case for $z=0$. 
The relative difference between the results from the \fr and \lcdm simulations are shown in the lower panels of Figure \ref{fig:bias_mass}. The simulations predict  $3-5\%$ lower bias in \fr compared to standard gravity at $z=0$. This relative difference is lower than the one found in lower resolution simulations  by \cite{schmidt2009}.  At redshift $z=1$ the difference in bias seems to depend more strongly on mass. At the low mass end the relative difference is the same as at $z=0$ while it drops to a $10\%$ lower bias for \fr compared to GR at the high mass end of the plot. 

\subsection{The halo concentration-mass relation}
The concentration mass relation for the dark matter halos at $z=0$ is shown in Figure \ref{fig:concentration}. The upper panels show the absolute value of the concentration  for all four simulations. 
These inferred from the circular velocity profile of the halos using
\citep{springel2008}
\begin{align}
\delta_c &= 7.213\, \delta_V = 7.213 \times 2 \left( \frac{v_{\rm max}}{H_0\, r_{\rm max}}\right)^2, \nonumber\\ \delta_c &= \frac{200}{3} \frac{c^3}{\log(1+c) - c / (1+c)},\label{concentration}
\end{align}
where $v_{\rm max}$ and $r_{\rm max}$ are the velocity and radius corresponding to the maximum of the profile and $c$ is the concentration parameter. 
We note that this method can lead to a weakly biased relative difference in the concentration between \fr and GR compared to profile fitting methods, particularly at the resolution limit of numerical simulations \citep{baldi2018}.
In addition to our simulation results we show theoretical fitting formulas from \cite{neto2007} and \cite{duffy2008} (our concentrations are calculated with respect to $r_{200\, {\rm crit}}$; we therefore choose the corresponding values for the fitting formula from \citealt{duffy2008}). In order to be consistent with the analysis in these papers we show the results for our full halo sample identified by \textsc{subfind} for each of the simulations (left panels) and for relaxed halos only (right panels). Our criteria for relaxed halos are the center-of-mass displacement and sub-mass criterion described in \cite{neto2007}. The center-of-mass displacement criterion limits the offset between a halos center-of-mass and its potential minimum to $0.07 \times r_{200\, crit}$ while the sub-mass criterion sets an upper bound of $10\%$ on the fraction of halo mass contained in substructures. 
The lower panels show the relative difference between \fr and a \lcdm universe.

As one can see from the upper panels, the \cite{neto2007} analytical formula provides an excellent fit to our standard gravity simulation results for both the relaxed and the full halo sample. The results for \fr match the fitting formula as well at the high mass end of the plot but show more concentrated profiles towards lower masses. At $10^{13.5}\ms/h$, the relative difference reaches $50\%$ for the full sample and $42\%$ for relaxed halos. This is expected as the stronger forces for unscreened (lower mass) objects in the modified gravity model move mass from the outer regions of the halo towards the center, leading to a steeper density profile \citep{arnold2016}. The increased concentration is also consistent with the results shown in \cite{schmidt2010} and \cite{shi2015}. The deviations of the results from the \lb simulation box towards low masses are likely caused by the limited resolution which makes it difficult to identify the maximum of the circular velocity profile.


\section{Summary and Conclusions}
\label{sec:conclusions}

We presented the to date (in terms of particle number) largest simulations of \cite{husa2007} \fr. The set of simulations we analysed consists of four simulations containing $2048^3$ simulation particles each,  in \ls  and \lb boxes for both \fr and a \lcdm model. Along with ordinary time-slice snapshots the simulations feature 2D and 3D lightcone outputs as well as FoF and \textsc{subfind} halo catalogs. We choose \fv as a background parameter for the scalar field for the modified gravity simulations.

Our findings can be summarised as follows:
\begin{itemize}
\item The matter power spectrum is increased in \fr on non-linear scales. The relative difference to GR is larger on smaller scales and grows with decreasing redshift. This result is consistent with previous works but extends to a much larger range in $k$. Comparing the power spectra of the two simulations with different resolution, we conclude that the standard and the modified gravity power spectra are affected in a very similar way at the resolution limit of the simulations which makes the relative differences between the different cosmological models trustworthy over a larger $k$-range compared to the absolute values of the individual spectra.
The growth of the BAO-oscillations is not affected by \fr. Differences between the gravity models appear only in the non-linear regime. 
Theoretical predictions for the non-linear matter power spectrum show good agreement with the simulations on large scales. They are nevertheless not accurate enough to precisely predict the relative difference between the cosmological models on smaller scales.
The angular power spectrum shows a -- within the Limber limit -- consistent behaviour.

\item The relative difference in the matter power spectrum between \fr and a \lcdm universe on small scales ($k\approx 10\hompc$) is of the same order of magnitude as the effect of baryonic processes such as feedback from AGN but acts in the opposite direction. Comparing our findings to results of the Eagle \citep{schaye2015} and Illustris TNG \citep{springel2018} hydrodynamical  simulations nevertheless suggests that there is a sweet spot around $k=1\hompc$ where the influence of baryons is very small but \fr has a sizeable effect on the power spectrum. We note that we can not make any statement about back-reactions between the two physical processes here. To make a conclusive statement about the interplay of baryonic physics and modified gravity it will be necessary to include both in one simulation at the same time. 

\item The changes to the linear and angular power spectrum are reflected in the lensing convergence spectrum. The relative difference in the lensing signal is again larger on smaller scales for the considered modified gravity model and reaches $25-30\%$ on the smallest scales probed by our simulations ($l\approx10^4$). Our simulation results match the predictions of \textsc{halofit} \citep{takahashi2012} and \textsc{mg-halofit} \citep{zhao2014} on large scales for the \lcdm model and \fr, respectively. On smaller scales the \mbox{\textsc{(mg-)halofit}} predictions overestimate our simulation results. 
A more detailed analysis of the lensing signal from the lightcone in \fr is planned in future work. 

\item The halo mass function is increased for intermediate mass halos by about $25\%$ in the considered modified gravity model. Halos at the high and low mass end of the correlation function are affected less. The position of the peak in the relative difference between the two gravity models depends on redshift. We observe the maximum relative difference around $10^{13} \ms$ at $z=1$ and around $10^{14} \ms$ at $z=0$.
The concentration mass relation is affected by \fr as well. While there is no significant difference between the cosmological models for masses above $10^{14.5} \ms$ the halos are more and more concentrated towards lower masses in the modified gravity simulations compared to their \lcdm counterparts. The relative difference reaches $50\%$ for the full halo sample and $40\%$ for relaxed halos at masses of $10^{13.5} \ms$. The results of our standard gravity simulations are in excellent agreement with the prediction of \cite{neto2007}.

\item The effects of \fr on the matter power spectrum is also reflected in the dark matter  auto-correlation function. Matter is more correlated on small scales in modified gravity. The relative differences reach $35\%$ at $r=10^{-2}{\rm Mpc}/h$.
In  contrast to matter, the dark matter halos are less correlated in modified gravity compared to GR. Independent of the mass of the halos considered  the halo-halo correlation function shows roughly $10\%$ lower values in \fr. 
The lower halo auto-correlation function results in a lower linear halo bias for modified gravity. Considering this bias as a function of mass we find that our GR results for $z=0$ and $z=1$ are in good agreement with the \cite{tinker2010} prediction while there is a clear difference to the \cite{sheth1999} model. The \fr simulations predict a lower bias for both redshifts compared to GR. At $z=1$ the relative difference is mass dependent and drops from $3\%$ at $10^{12}\ms$ to $10\%$ at $10^{13.5} \ms$. Our $z=0$ result does not show a clear mass dependence. The simulations predict about $3-5\%$ lower halo bias for this redshift. It is however worth noting that the difference between the models is significantly smaller than the difference between the different theoretical predictions for a \lcdm universe.

\end{itemize}

All in all we conclude that the modified gravity lightcone simulation suite provides high resolution, large volume simulation data in \fr which allows to analyse the effect of modified gravity onto cosmic structure formation over a range of scales unreached so far. 
The high resolution lightcone simulations presented in this paper are a valuable tool for exploring possible deviations of modified gravity models with respect to LCDM for a wide range of observables. Galaxy mocks based on this set of simulations and their properties will be presented in a forthcoming publication.
The results presented in this paper show that the simulations are consistent with previous works and theoretical expectations
and show their robustness against mass-resolution effects, indicating that these simulations can be safely used to test gravity using the large-scale distribution of matter and galaxies.

\section*{Acknowledgements}
The authors like to thank Marco Baldi, Kazuya Koyama, Claudio Llinares and Baojiu Li for useful discussions and comments. 
Some of the results in this paper have been derived using the \textsc{healp}ix \citep{healpix}  package.

CA acknowledges support from the European Research Council  through ERC-StG-716532-PUNCA. 
PF acknowledges support from MINECO through grant ESP2015-66861-C3-1-R,
and Generalitat de Catalunya through grant 2017-SGR-885.
VS acknowledges support from the Deutsche Forschungsgemeinschaft (DFG) through Transregio 33,  ''The Dark Universe''.  
EP acknowledges support by the Kavli Foundation.
L.B. acknowledge the support from the Spanish Ministerio de Economia y Competitividad grant ESP2015-66861.

The simulations performed for this work were run on the Jureca cluster at the Juelich Supercomputing Center in Juelich, Germany within project HHD29, on Hazelhen at the High-Performance Computing Center Stuttgart in Stuttgart and on the bwForCluster MLS\&WISO Developement. 
This work used the DiRAC Data Centric system at Durham University, operated by the Institute for Computational Cosmology on behalf of the STFC DiRACHPC Facility (www.dirac.ac.uk). This equipment was funded by BIS National E-infrastructure capital grant ST/K00042X/1, STFC capital grants ST/H008519/1 and ST/K00087X/1, STFC DiRAC Operations grant ST/K003267/1 and Durham University. DiRAC is part of the National E-Infrastructure.

\bibliographystyle{mnras}
\bibliography{paper}


\end{document}